\documentclass[10pt,preprint]{aastex}
\pdfoutput=1

\newcommand*{\teff}{$T_{\rm eff}$}
\newcommand*{\logg}{$\log~g$}
\newcommand*{\feh}{[Fe/H]}
\newcommand*{\tcrit}{$T_{\rm crit}$}
\newcommand*{\afe}{[$\alpha$/Fe]}
\newcommand*{\kms}{km s$^{-1}$}
\newcommand*{\zmax}{$Z_{\rm max}$}

\newcommand*{\vphi}{$V_{\rm \phi}$}

\newcommand*{\z}{$|Z|$}

\newcommand*{\Sch}{Sch\"onrich et al.}

\usepackage{graphicx}

\shorttitle{The Case for the Dual Halo of the Milky Way}
\shortauthors{Beers et al.}

\begin{document}

\title{The Case for the Dual Halo of the Milky Way}

\author{Timothy C. Beers\altaffilmark{1,2},
Daniela Carollo\altaffilmark{3,4},
\v Zeljko Ivezi\'c\altaffilmark{5},
Deokkeun An\altaffilmark{6},
Masashi Chiba\altaffilmark{7},
John E. Norris\altaffilmark{3},
Ken C. Freeman\altaffilmark{3},
Young Sun Lee\altaffilmark{1},
Jeffrey A. Munn\altaffilmark{8},
Paola Re Fiorentin\altaffilmark{9},
Thirupathi Sivarani\altaffilmark{10},
Ronald Wilhelm\altaffilmark{11},
Brian Yanny\altaffilmark{12},
Donald G. York\altaffilmark{13}}

\altaffiltext{1}{Department of Physics \& Astronomy and JINA (Joint Institute for Nuclear Astrophysics),
                 Michigan State University, East Lansing, MI 48824, USA;
                 beers@pa.msu.edu; lee@pa.msu.edu }

\altaffiltext{2}{National Optical Astronomical Observatory, Tucson, AZ 85719, USA}

\altaffiltext{3}{Research School of Astronomy \& Astrophysics, Australian
National University, Mount Stromlo Observatory, Cotter Road, Weston, ACT, 2611,
Australia; carollo@mso.anu.edu.au, jen@mso.anu.edu.au; kcf@mso.anu.edu.au}

\altaffiltext{4}{Department of Physics \& Astronomy, Macquarie University, Sydney, NSW, 2109, Australia}

\altaffiltext{5}{Department of Astronomy, University of Washington, Box 351580, Seattle, WA 98195, USA; ivezic@astro.washington.edu}

\altaffiltext{6}{Department of Science Education, Ewha Womans University, Seoul 120-750, Republic of Korea; deokkeun@ewha.ac.kr}

\altaffiltext{7}{Astronomical Institute, Tohoku University, Sendai 980-8578, Japan; chiba@astr.tohoku.ac.jp}

\altaffiltext{8}{U.S. Naval Observatory, Flagstaff Station, 10391 W. Naval Observatory Road, Flagstaff, AZ 86001, USA; jam@nofs.navy.mil}

\altaffiltext{9}{INAF-Osservatorio Astronomico di Torino, via Osservatorio 20, 10025 Pino Torinese, Italy; refiorentin@oato.inaf.it}

\altaffiltext{10}{Indian Institute of Astrophysics, II Block, Koramangala, Bangalore 560 034, India; sivarani@iiap.res.in}

\altaffiltext{11}{Physics and Astronomy Department, University of Kentucky, Lexington, KY 40506; rjwi222@uky.edu}

\altaffiltext{12}{Fermi National Accelerator Laboratory, Batavia, IL 60510, USA; yanny@fnal.gov}

\altaffiltext{13}{Department of Astronomy and Astrophysics, University of Chicago, Chicago, IL 60637, USA; don@oddjob.uchicago.edu}

\begin{abstract}

Carollo et al. have recently resolved the stellar population of the Milky
Way halo into at least two distinct components, an inner halo and an outer
halo. This result has been criticized by Sch\"onrich et al., who claim that
the retrograde signature associated with the outer halo is due to the
adoption of faulty distances. We refute this claim, and demonstrate that
the \Sch\ photometric distances are themselves flawed because {\it they
adopted an incorrect main-sequence absolute magnitude relationship from the
work of Ivezi\'c et al.}. When compared to the recommended relation from
Ivezi\'c et al., which is tied to a Milky Way globular cluster distance
scale and accounts for age and metallicity effects, the relation adopted by
\Sch\ yields up to 18\% shorter distances for stars near the main-sequence turnoff (TO). 
Use of the correct relationship yields agreement between the distances
assigned by Carollo et al. and Ivezi\'{c} et al. for low-metallicity dwarfs
to within 6-10\%. \Sch\ also point out that intermediate-gravity stars (3.5
$\le \log g <$ 4.0) with colors redder than the TO region are likely
misclassified, with which we concur. We implement a new procedure to
reassign luminosity classifications for the TO stars that require it. New
derivations of the rotational behavior demonstrate that the retrograde
signature and high velocity dispersion of the outer-halo population
remains. We summarize additional lines of evidence for a dual halo,
including a test of the retrograde signature based on proper motions alone,
and conclude that the preponderance of evidence strongly rejects the
single-halo interpretation.

\end{abstract}


\section{Introduction}

The nature of the stellar halo of the Milky Way has been debated for many
decades. Among the questions that have been asked: Is the halo a monolithic
structure, well-described by a simple Gaussian velocity ellipsoid? If so,
is it in zero net rotation, and does that rotational character apply to all
of its constituent stars? Do the stars in the halo comprise a single
stellar population, with similar ages and drawn from a common metallicity
distribution function (MDF)? Can the spatial distribution of the halo stars
be adequately described by a single density law (power-law or otherwise)?
Due to the difficulty of teasing out the properties of such a low-density
component (as compared, e.g., to the bulge and disk systems), the basic
data required to address these and other questions has only recently begun
to arrive. Not surprisingly, multiple interpretations have emerged.

Massive new datasets from, e.g., SkyMapper (Keller et al. 2007), Gaia
(Perryman et al. 2001), and eventually, LSST (Ivezi\'{c} et al. 2008a),
will provide definitive answers to the above questions, and of course,
raise new ones. However, it is critical to address these issues with
presently available data, so that the most meaningful probes of future
datasets can be developed.

The two largest spectroscopic datasets available today for examination of
the stellar populations of the Milky Way are the RAdial Velocity Experiment
(RAVE; Steinmetz et al. 2006; Zwitter et al. 2008; Siebert et al. 2011) and
the Sloan Digital Sky Survey (SDSS; York et al. 2000), in particular the
subsurvey Sloan Extension for Galactic Understanding and Exploration
(SEGUE; Yanny et al. 2009). The SEGUE-2 subsurvey (Rockosi et al., in
preparation) has recently been publicly released as part of SDSS DR8
(Aihara et al. 2011), and will add to this bounty of information. For now,
we concentrate on the information available from the previous public
release from SDSS, DR7 (Abazajian et al. 2009), and in particular address
the criticisms raised by \Sch\ (2010; S10) of the previous work of Carollo
et al. (2007; C07) and Carollo et al. (2010; C10).  

Note that S10 is the version of the \Sch\ manuscript that appeared as
arXiv:1012.0842v1. In their published paper (\Sch\ 2011; S11), these
authors chose to respond to the submitted version of the present paper,
which appeared as arXiv:1104.2513v1. Due to the potential confusion over the
issues raised in the two versions of the \Sch\ drafts, we have confined our
analysis below to the version that appeared as S10; in the Appendix we
briefly consider the issues raised by S11.

Carollo et al. (2007) performed a kinematic analysis (within a local
volume) for a large sample of calibration stars from SDSS DR5
(Adelman-McCarthy et al. 2007), and argued for the existence of at least a
two-component halo. In their view the Galactic halo comprises two broadly
overlapping structural components, an inner halo and an outer halo. Note
that these labels are not merely descriptors for the regions studied, but
rather are labels for two individual stellar populations. These components
exhibit different spatial-density profiles, stellar orbits, and stellar
metallicities. It was found that the inner-halo component dominates the
population of halo stars found at distances up to 10-15 kpc from the
Galactic center, while the outer-halo component dominates in the region
beyond 15-20 kpc. The inner halo was shown to comprise a population of
stars exhibiting a flattened spatial density distribution, with an inferred
axial ratio on the order of $\sim$ 0.6. According to C07, inner-halo stars
possess generally high orbital eccentricities, and exhibit a small (or
zero) net prograde rotation around the center of the Galaxy. The MDF of the
inner halo peaks at [Fe/H] $= -1.6$, with tails extending to higher and
lower metallicities. By comparison, the outer halo comprises stars that
exhibit a more spherical spatial-density distribution, with an axial ratio
$\sim$ 0.9. Outer-halo stars possess a wide range of orbital
eccentricities, exhibit a clear retrograde net rotation, and are drawn from
an MDF that peaks at [Fe/H] $= -2.2$, a factor of four lower than that of
the inner-halo population.

Carollo et al. (2010) used an expanded sample of calibration stars
available from SDSS DR7, which included the SEGUE sample, to refine and
extend the results of C07. They derived velocity ellipsoids for the inner-
and outer-halo components of the Galaxy, as well as for the canonical
thick-disk and the proposed metal-weak thick-disk populations. The C10
paper also considered the fractions of each component required to
understand the nature of the observed kinematic behavior of the stellar
populations of the Galaxy as a function of distance from the Galactic
plane. Spatial-density profiles for the inner- and outer-halo populations
were inferred from a Jeans Theorem analysis. The full set of calibration
stars (including those outside the local volume) was used to test for the
expected changes in the observed stellar MDF with distance above the
Galactic plane {\it in situ}, due to the changing contributions from the
underlying stellar populations.

Derivation of sufficiently accurate distances is a crucial required step in
carrying out kinematic analyses that make use of full space motions, as
these involve distances, combined with radial velocities and proper
motions, in order to assemble the local velocity components of a sample. It
is these distances that have been called into question by S10. In the
present paper, we show that many of their objections arise from {\it their
incorrect adoption} of a main-sequence absolute magnitude relationship from
Ivezi\'c et al. (2008b; I08) that {\it does not apply} for metal-poor halo
stars near the main-sequence turnoff (TO), and which leads to assignments
of stellar distances that strongly disagree (a shorter scale by 10-18\%)
with those derived using the correct relationship recommended by I08. A
legitimate criticism by S10 relates to the luminosity classifications for
stars of intermediate gravity (as assigned spectroscopically) used by C07
and C10, which we demonstrate below is easily corrected. We then consider a
new kinematic analysis of likely outer-halo stars from C10, and demonstrate
that their original claim that the halo of the Milky Way requires at least
a two-component model (with the outer-halo component in net retrograde
rotation and possessing a large velocity dispersion) remains intact.

This paper is outlined as follows. In Section 2 we summarize the procedures
used by C07 and C10 to derive absolute magnitudes and distance estimates
for their stars, which were based on those described by Beers et al. (2000)
. A technique for the reassignment of (some of the) luminosity
classifications for TO stars in the original C10 sample is then developed
and applied. In Section 3 we compare with absolute magnitudes and distances
derived by the approaches of I08 and An et al. 2011 (in preparation; A11)
for stars spectroscopically classified as likely dwarfs based on their
derived surface gravities, as well as with those claimed by S10. We
demonstrate concordance between the distances for low-metallicity dwarf
stars obtained by C10, I08, and A11, and the apparent discordance of all
three of these techniques with the results of S10. In Section 4 we
reanalyze the kinematics of likely outer-halo stars from the C10 dwarf
sample, as well as from the full sample, including stars of dwarf, TO, and
subgiant/giant luminosity classifications, and compare to the results
obtained from adoption of the I08, A11, and S10 distances. Additional tests
for the presence of a kinematically and/or chemically distinct outer halo
in the C10 sample are discussed in Section 5. Section 6 presents a summary
of further evidence in favor of a dual halo model for the Milky Way, based
on other data sets from SDSS and elsewhere. Our conclusions are given in
Section 7. In the Appendix, we consider the issues raised by S11 (the
published version of S10).

\section{Procedures Used for Absolute Magnitude and Distance Estimates}

\subsection{As Employed by C07 and C10}

The analyses of C07 and C10 made use of distance estimates for various
luminosity classes as assigned by the software pipeline employed by
SDSS/SEGUE to estimate stellar atmospheric parameters based on
low-resolution ($R \sim 2000$) spectroscopy and $ugriz$ photometry. The
SEGUE Stellar Parameter Pipeline (SSPP) assigns distances for stars under
the following assumed luminosity classes -- D: dwarf, TO: main-sequence
turnoff, SG/G: subgiant and giant, FHB: Field Horizontal-Branch, and AGB:
Asymptotic Giant Branch.\footnote{The FHB and AGB classes do not pertain to
the sample of calibration stars used by C07 and C10, and so are not
discussed further here.} Details of the development, calibration, and
validation of the SSPP can be found in Lee et al. (2008a,b), Allende Prieto
et al. (2008), and Smolinski et al. (2011), to which we refer the
interested reader.

The SSPP obtains estimates of stellar effective temperatures, \teff, with
errors of determination on the order of 150~K. The surface gravity
estimates returned by the SSPP are accurate, for stars other than the
coolest giants, to on the order of 0.25 dex. Metallicity estimates for
stars in the temperature range 4500~K $< T_{\rm eff} < 7000$~K are accurate
to on the order of 0.2 dex.

The SSPP distance estimates for various luminosity classes are based on a
set of absolute magnitude relationships (using absorption and
reddening-corrected Johnson $V$ magnitudes and $B-V$ colors) calibrated to
Galactic globular and open clusters, as described by Beers et al. (2000;
their Table 2). As demonstrated in Beers et al. (2000), photometric
distances estimated for their sample are in good agreement with distances
derived from accurate Hipparcos parallaxes. Even when confined to TO stars
alone (with well-examined assignment of stars into the TO class provided
from previous work), the photometric distances using the Beers et al.
formulae are consistent with Hipparcos distances.

The samples used by C07 and C10 were selected from the calibration stars of
SDSS/SEGUE, which cover an apparent magnitude range of $15.5 < g_0 < 18.5$.
In those analyses, confinement to a local sample with distances less than 4
kpc from the Sun corresponds to a $g$-band absolute magnitude fainter than
$M_g = 2.5$, i.e., the local sample is dominated by D and TO stars. This is
in contrast to the sample considered by Beers et al. (2000), which is
dominated by SG/G stars.

Since the Beers et al. (2000) approach makes use of a non-SDSS photometric
system, it is also necessary to employ a color transformation from the SDSS
system. Zhao \& Newberg (2006) derived a transformation obtained by making
matches of SDSS stars with available Johnson magnitudes and colors from the
HK survey of Beers and colleagues (Beers et al. 1985, 1992), as well as
additional photometry of the HK sample stars obtained over the past decade
(see, e.g., Beers et al. 2007, and references therein). They obtained:

$$V = g - 0.561\,(g-r)- 0.004$$
$$B-V = 0.916\,(g-r) + 0.187$$

Stars from the HK survey were used in order to specifically include stars
with [Fe/H] $< -1.0$, which pertain to most halo stars, although the
results did not differ drastically from those of Fukugita et al. (1996)
that were based primarily on higher abundance stars. The color range of the
matching stars sets the region of applicability of the above
transformation, which is $-0.5 < g-r < 1.0$. The choice of distance
estimates based on a non-SDSS photometric system was one of necessity at
the time the SSPP was put into operation, as there were no suitably
calibrated fiducials based on SDSS photometry of Galactic clusters
available, and the isochrones that had been developed were rather
primitive. These limitations no longer apply, and future versions of the
SSPP will employ alternative distance estimates based on improvements that
have become available in the past year.

It should be noted that the SSPP, by design, does not identify a preferred
distance estimate, leaving the choice of the appropriate luminosity
classification to the user's discretion. This choice is due, in part, to
the fact that the estimation of surface gravity by the SSPP has evolved
with time, and may continue to do so in the future. Hence, as many users
will rely, at least at some level, on \logg\ estimates for making distance
estimates based on luminosity classifications from available spectroscopic
information, no ``approved'' distance estimate is supplied by the SSPP.

For the purpose of the analyses carried out by C07 and C10, the following
spectroscopically assigned surface gravity intervals from the SSPP were
used in the assignment of luminosity classifications:

\begin{itemize}

\item D: \logg\ $\ge 4.0$

\item TO:  $3.5 \le$ \logg\ $< 4.0$

\item SG/G:  \logg\ $< 3.5$

\end{itemize}

Estimates of \logg\ carry errors, and one has to be concerned about the
possible effects on any resulting analyses based on their adoption. For the
present, this is best assessed by consideration of inferences based on
samples of individual luminosity classes relative to the sample as a whole,
which we discuss below.

Note that the above prescription for assignment of luminosity class does
not take into account the ``known'' evolutionary stage of a given star, as
might be inferred from the location of a star in a color-magnitude diagram
expected to pertain to objects of a given age and metallicity. This
uncertainty is of particular concern for stars assigned to the TO class,
since an alternative assignment to the D or SG/G class could result in
potentially large discrepancies in the adopted distance. This ``defect''
(actually a choice, given that such knowledge is at best only partially
constrained with present data, and in any case relies on assumptions
regarding the underlying stellar population one adopts) is one of the
criticisms of the C07 and C10 work levied by S10. However, it can be
readily addressed, as described below.

As part of their analysis, I08 compared absolute magnitude estimates
obtained by the Beers et al. (2000) procedures with those used in their own
analysis (which only applied to dwarfs). Pointing at the bottom left panel
of their Fig. 21, which examined the main-sequence comparisons of Galactic
clusters between the two studies, I08 concluded that ``... the median
offset of implied $M_r$ evaluated in small bins of $u-g$ and $g-r$ color is
$-0.07$ mag, with an rms of 0.06 mag''. This satisfying level of agreement
provided additional reason to have faith in the distances for the majority
of stars in the C10 sample upon which their kinematic analysis was based.
This agreement remains intact, as shown below.

\subsection{A Refined Prescription for Luminosity Class Assignments}

As pointed out above, refinements in luminosity class assignment require
assumptions about the ages and age distributions of the population(s) to
which they will be applied. For the present discussion, which turns on the
nature of the stars associated by C07 and C10 with the inner- and
outer-halo populations, it is reasonable to adopt a uniformly old age, with
the unavoidable caveat that not all stars of these populations may strictly
adhere to this assumption.

We proceed as follows.

First, a set of theoretical \logg\ vs. \teff\ diagrams is obtained, based
on the Y$^2$ isochrones (Demarque et al. 2004), for a population with age
set to 12 Gyrs, metallicities in the range $-3.0 \le$ [Fe/H] $ \le 0.0$,
and with \afe\ set to 0.0 for solar metallicity, \afe\ = +0.3 for \feh\
$\le -1.0$, and using a linear scaling between [Fe/H] = 0 and \feh\ $=
-1.0$. We then obtain the effective temperatures at the position of the
main-sequence turnoff for each model, $T_{\rm MSTO}$, and assign a
``critical temperature'', \tcrit, to be 250~K cooler than $T_{\rm MSTO}$.
The offset of 250~K was chosen since, in the region of the MSTO, this
roughly corresponds to the two-sigma accuracy of the estimated temperature
from the SSPP, and provides a reasonable location for the base of the
subgiant branch for isochrones of old, low-metallicity populations. Our
purpose is to define a criterion such that a reassignment of luminosity
classes can be considered for stars of intermediate gravity ($3.5
\le \log g < 4.0$) that are cooler than \tcrit.

A second-order polynomial is then fit to the positions of the $T_{\rm
MSTO}$ values for each model:

\begin{equation}
T_{\rm MSTO} = 5572 - 519.3\,{\rm [Fe/H]} - 44.3\,{\rm [Fe/H]}^2
\end{equation}

\begin{figure}
\hspace{-1cm}
\epsscale{0.75}
\plotone{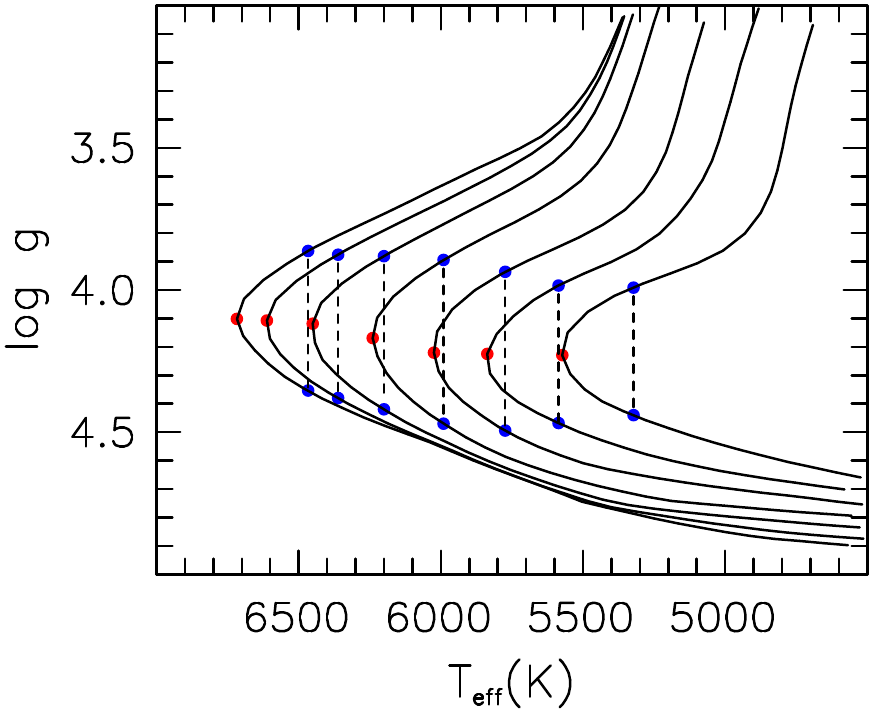}
\caption{Theoretical surface gravity ($\log g$) vs. \teff\ diagram based on the Y$^2$
isochrones (Demarque et al. 2004), under the assumption of a uniform age of 12
Gyrs. From left to right in the figure, the isochrones cover the range of
metallicities $-3.0 < $ [Fe/H] $< 0.0$, in steps of 0.5 dex. The \afe\ ratios
are set to 0.0 for solar metallicity, \afe\ = +0.3 for \feh\ $\le -1.0$, and are
linearly scaling between [Fe/H] = 0 and \feh\ $= -1.0$. The red dots mark the
position of the MSTO, while the blue dots correspond to temperatures 250~K
cooler than the MSTO, referred to as \tcrit. The vertical dashed lines
connect the multi-valued positions on the isochrones at a given \teff. Stars
with estimated gravities in the range $3.5 \le$ \logg\ $< 4.0$, which were
classified as TO by C07 and C10, are reassigned to either D or SG/G classes if
their metallicities and \teff\ place them to the right side of these divisions,
or remain classified as TO if their metallicities and \teff\ place them to the
left of these divisions. See text for additional details.}
\label{fig:tcrit_plot}
\end{figure}

\medskip
\noindent The critical temperature is then simply set to \tcrit\ = $T_{\rm MSTO} -
250$~K. This process is illustrated in Fig.~\ref{fig:tcrit_plot}. The
critical temperature is used in order to separate intermediate-gravity
stars classified as TO by C07 and C10 into either bona-fide TO stars (those
with \teff\ $\ge$ \tcrit\ ) or into the D or SG/G classes (those with
\teff\ $<$ \tcrit\ ) according to their surface gravity estimates, as
summarized in Table 1. Note that stars with original luminosity
classifications D and SG/G are not changed by this procedure.

\begin{figure*}
\centering
\includegraphics[angle=90,scale=0.60]{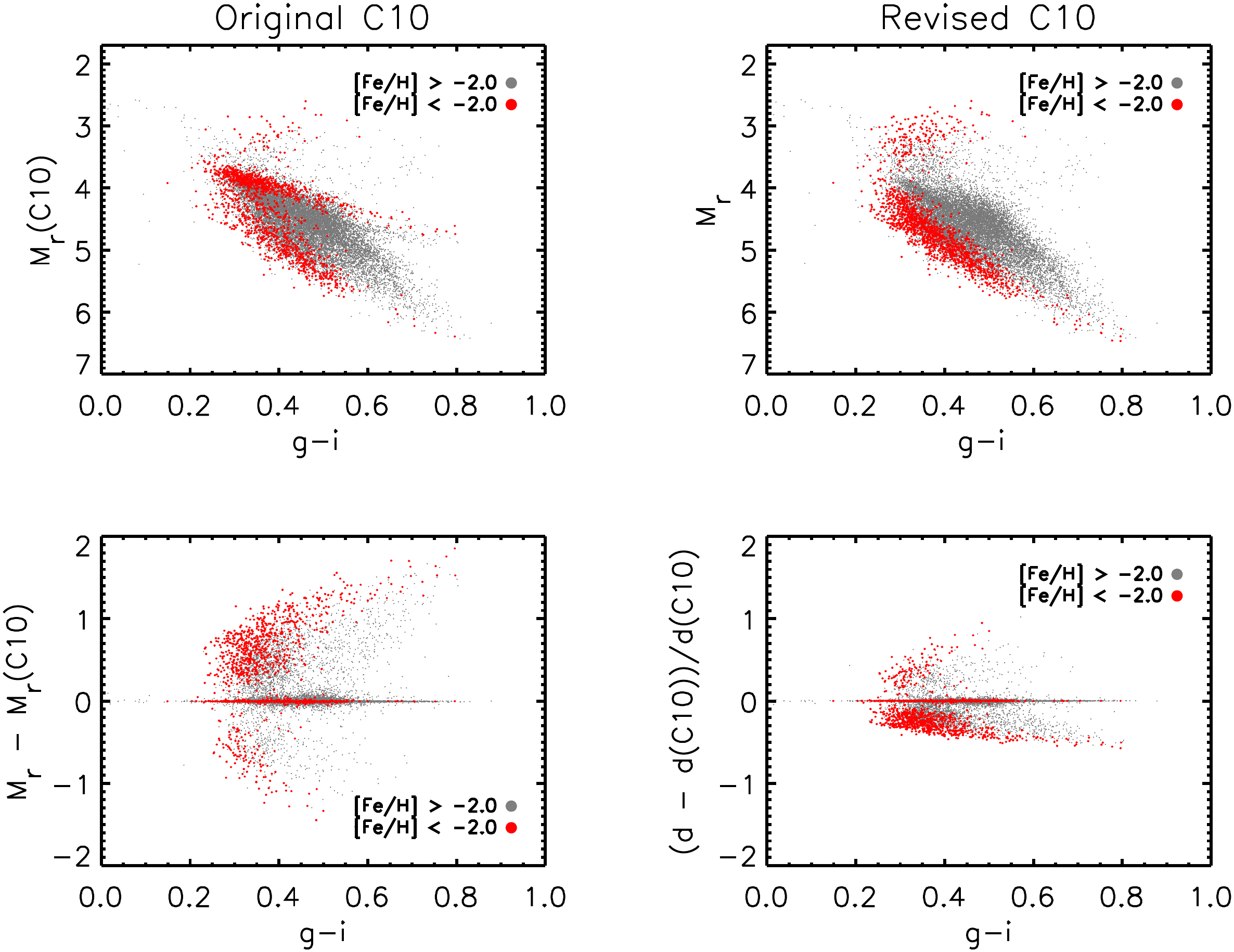}
\caption{{\it Upper left:} $M_r, g-i$ CMD for the C10 sample,
using the original luminosity classifications. {\it Upper right}: $M_r, g-i$ CMD
for the C10 sample, using the revised luminosity classifications, as described
in the text. The stars with [Fe/H] $ > -2.0$ are shown as gray dots, while those
with [Fe/H] $ < -2.0$ are shown as red dots. {\it Lower left}: Difference
between the $M_r$ absolute magnitudes for the revised luminosity classifications
and the original C10 classifications, as a function of $g-i$. {\it Lower right}:
Fractional change in derived distances for the revised luminosity
classifications vs. the original C10 classifications, as a function of $g-i$.}
\label{fig:mags_dist}
\end{figure*}

The luminosity class reassignment procedure described above affects a total
of 4514 of the original 16920 accepted stars in the C10 sample (26\%). The
upper left panel of Fig.~\ref{fig:mags_dist} shows the CMD for the original
assignments of C10, while the upper right panel is that obtained after the
revised assignments have been applied to this same sample. The gray dots
are stars with [Fe/H] $ > -2.0$, while the red dots are stars with [Fe/H] $
< -2.0$.\footnote{We have made use of the corrected metallicity,
[Fe/H]$_C$, as described by C10, here and throughout the rest of this
paper, for the quoted metallicities.} As can be appreciated from comparison
of these two panels, stars that formerly fell into regions of the CMD that
might be considered astrophysically unlikely for an old, metal-poor
population have primarily moved into either the D or SG regions. The lower
panels of Fig.~\ref{fig:mags_dist} contrast the absolute magnitudes of the
revised and original C10 classifications (lower left) and the corresponding
derived distances (lower right).

Inspection of the upper left panel of Fig.~\ref{fig:mags_dist} clearly
shows the presence of the ``spurious'' TO stars in the original C10 sample,
most easily seen among the [Fe/H] $< -2.0$ stars as the plume extending
from roughly $M_r = 3.7$ to $M_r = 4.7$, over the color range $0.25 < g-i <
0.6$. Comparison with the upper right panel of this figure shows that most
of these stars (51\%) are reassigned to D status, with only some 10\% being
reassigned to SG/G status (the remaining stars, 39\%, retain their original
luminosity classification of TO). At low metallicity, [Fe/H] $< -$2.0, the
fraction of reassigned TO stars to D status is 85\%, while those reassigned
to SG/G status comprise 14\%, and only a small fraction retain their TO
classification. At higher metallicities, [Fe/H] $> -$2.0, 44\% of the TO
stars are reassigned to D status, and only a small fraction are reassigned
to SG/G status. The remaining stars, 56\%, retain their original luminosity
classification of TO.

The lower left panel of Fig.~\ref{fig:mags_dist} shows the difference in
the assigned $M_r$ absolute magnitudes that arises when one compares the
revised C10 estimates with those of C10. For the TO stars that were
reclassified as D stars, and with [Fe/H] $> -2.0$, the revised C10
determinations are fainter by a median offset of 0.08 mags (rms 0.36 mags)
for $0.4 < g-i < 0.8$, while the median offset of the revised C10 absolute
magnitudes is 0.30 mags (rms 0.24 mags) fainter for bluer stars in the
range $g-i < 0.4$. For the TO stars that were reclassified as SG/G stars,
and with [Fe/H] $> -2.0$, the revised C10 determinations are brighter by a
median offset of 0.48 mags (rms 0.31 mags) for $0.4 < g-i < 0.8$, while the
median offset of revised C10 absolute magnitudes is 0.44 mags (rms 0.22
mags) brighter for bluer stars in the range $g-i < 0.4$.

For the TO stars that were reclassified as D stars, and with [Fe/H] $<
-2.0$, the revised C10 determinations are fainter by a median offset of
0.97 mags (rms 0.43 mags) for $0.4 < g-i < 0.8$, while the median offset of
revised C10 absolute magnitudes is 0.60 mags (rms 0.25 mags) fainter for
bluer stars in the range $g-i < 0.4$. For the TO stars that were
reclassified as SG/G stars, and with [Fe/H] $< -2.0$, the revised C10
determinations are brighter by a median offset of 1.07 mags (rms 0.42 mags)
for $0.4 < g-i < 0.8$, while the median offset of revised C10 absolute
magnitudes is 0.63 mags (rms 0.24 mags) brighter for bluer stars in the
range $g-i < 0.4$. 

The lower right panel of Fig.~\ref{fig:mags_dist} shows the fractional
difference in the derived distances between the revised C10 and C10 scales.
For TO stars that were reclassified as D stars, and with [Fe/H] $> -2.0$
and $0.4 < g-i < 0.8$, the median offset of the revised C10 distances with
respect to the C10 distances is 26\% (rms 9\%). In the bluer range, $g-i <
0.4$, the offset increases to about 19\% (rms 6\%). Both revisions are in
the direction that the revised C10 scale is shorter than the original C10
scale for the reclassified TO $\rightarrow$ D stars. For TO stars that were
reclassified as SG/G stars, and with [Fe/H] $> -2.0$ and $0.4 < g-i < 0.8$,
the median offset of the revised C10 distances with respect to the C10
distances is 33\% (rms 16\%). In the bluer range, $g-i < 0.4$, the offset
decreases to about 25\% (rms 11\%). Both revisions are in the direction
that the revised C10 scale is longer than the original C10 scale for the
reclassified TO $\rightarrow$ SG/G stars.

For TO stars that were reclassified as D stars, and with [Fe/H] $< -2.0$
and $0.4 < g-i < 0.8$, the median offset of the revised C10 distances with
respect to the C10 distances is 36\% (rms 14\%). In the bluer range, $g-i <
0.4$, the offset decreases to about 24\% (rms 9\%). Both revisions are in
the direction that the revised C10 scale is shorter than the original C10
scale for the reclassified TO $\rightarrow$ D stars. For TO stars that were
reclassified as SG/G stars, and with [Fe/H] $< -2.0$ and $0.4 < g-i < 0.8$,
the median offset of the revised C10 distances with respect to the C10
distances is 65\% (rms 26\%). In the bluer range, $g-i < 0.4$, the offset
decreases to about 34\% (rms 14\%). Both revisions are in the direction
that the revised C10 scale is longer than the original C10 scale for the
reclassified TO $\rightarrow$ SG/G stars.

It is worth considering that our reclassification procedure {\it assumes}
that many of the stars with spectroscopically assigned surface gravities in
the range $3.75 \le \log g < 4.0$ (those significantly cooler than an
inferred old-population main-sequence turnoff) are indeed metal-poor dwarfs
with slightly misestimated \logg. This is certainly a conservative
assumption, and errs on the side of decreasing distances for actual TO or
SG/G stars to the much smaller values that would be derived if they are in
fact main-sequence dwarfs. These fine adjustments require further study and
verification by high-resolution spectroscopic follow-up of a sample of such
stars, at a variety of metallicities and temperatures.

Note that for construction of Figures 2-8, and for the distance-scale
comparisons we carry out below, it is useful to consider samples that
explore the same local volumes. For simplicity, and for consistency with
C07 and C10, we have selected stars with revised C10 distance estimates
that satisfy $7 < R < 10$ kpc and $d < 4$ kpc as our basis sample.

\subsection{Comparison Between Revised C10 and S10}

The essence of the S10 complaint is that the distance scale utilized by C07
and C10 is too ``long'', i.e., that we have artificially inflated the
estimates of stellar distances through the combination of (1) the use of
misclassified TO stars (which they suggest could be D stars instead), and
in particular, (2) the use of an absolute magnitude scale for the D stars
that assigns luminosities to main-sequence stars which displaces them to
larger-than-appropriate distances. We have shown above that the first issue
is easily corrected for, and that in any case it only applies to some 14\%
of the total calibration stars from C10, roughly 2300 stars. Of these, 4\%
of the full sample (680 stars) possess the very low metalliticties (below
[Fe/H] $= -2.0$) that strongly influence the derived properties of a
proposed outer-halo population. Thus, even if there might be some impact,
it is substantially diluted by the relatively small numbers of stars for
which this concern exists. In any case, we have applied the correction
procedures described above, carried out the luminosity classification
changes for the cooler TO stars, and in the analysis below, refer to the
modified sample as the revised C10 sample. The second issue turns on
whether or not one should put faith in our adopted main-sequence absolute
magnitude scale, which we address in detail below.

\begin{figure*}
\hspace{-4.5cm}
\includegraphics[angle=90,scale=0.75]{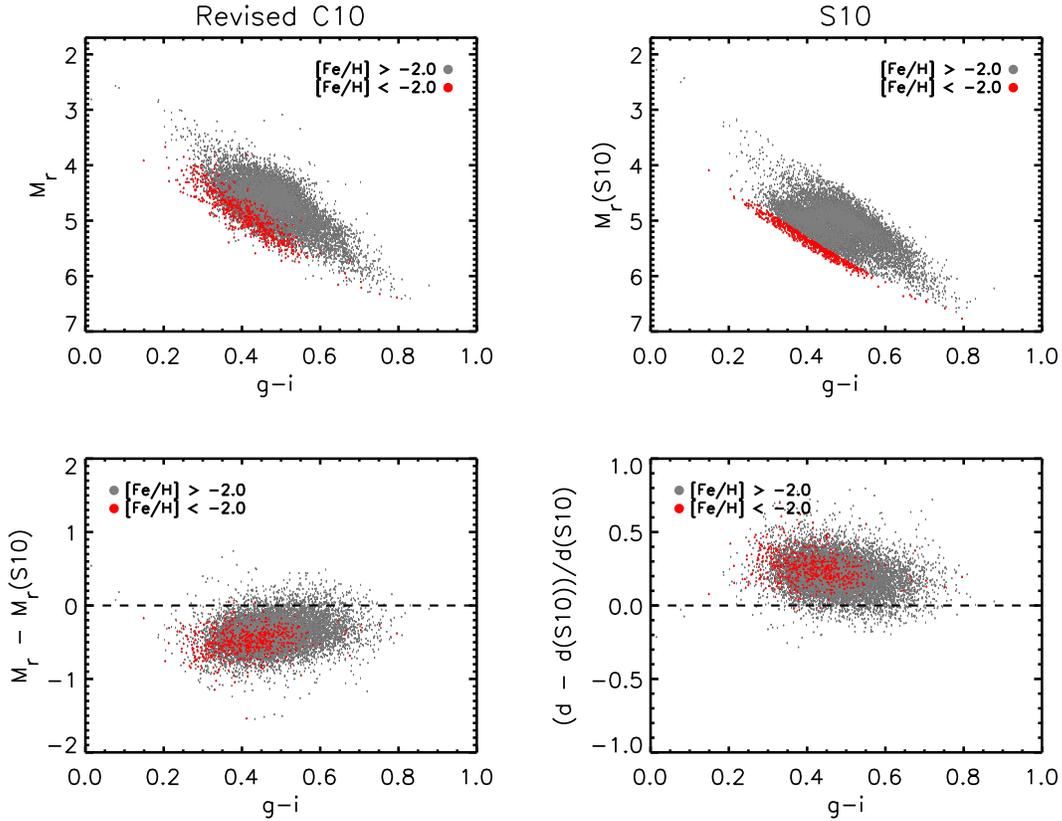}
\caption{{\it Upper left}: M$_r, g-i$ CMD for the revised C10 luminosity
classifications and with spectroscopically assigned D classifications. {\it
Upper right}: M$_r, g-i$ CMD for stars with spectroscopically assigned D
classifications, with absolute magnitudes calculated from Eqn. (A1) of I08, as
adopted by S10. The stars with [Fe/H] $> -2.0$ are shown as gray dots, while
those with [Fe/H] $< -2.0$ are shown as red dots. {\it Lower left}: Difference
between the M$_r$ absolute magnitudes for stars with spectroscopically assigned
D classifications for the revised C10 and S10 calculations, as a function of
$g-i$. {\it Lower right}: Fractional change in derived distances from the
revised C10 sample as compared to those adopted by S10, as a function of $g-i$.}
\label{fig:mag_comp_S10}
\end{figure*}

Fig.~\ref{fig:mag_comp_S10} shows the result of the comparison of the
revised C10 determinations with those of S10. The upper left panel of this
figure shows the CMD for stars with spectroscopic assignments of D ($\log g
\geq 4.0$), with absolute magnitudes from the revised C10 sample. The upper
right panel shows the corresponding CMD obtained using the absolute
magnitudes from S10 (Eqn. 3 below). Note that in the evaluation of both
relationships, the [Fe/H]$_C$ from C10 was employed, although similar
results are obtained when the adopted metallicities from the SSPP
([Fe/H]$_A$) are used. The stars are color-coded to indicate metallicities
above and below [Fe/H] $ = -2.0$.

Note that S10 did make a number of changes in their adopted absolute
magnitude relationship relative to Eqn. (A1) of I08, which actually serve
to bring their estimated distances into closer agreement with ours. We have
not attempted to recreate these adjustments in our analysis, as the
corrections they apply are themselves uncertain (and in our view not
entirely well-motivated, e.g., their preference for metallicities on a
scale that our own analysis does not support). Thus, one should properly
consider the comparisons we make here as likely to be the maximum
differences that would be obtained. The apparent difference in the scatter
in absolute magnitudes seen in the upper left and upper right panels is due
to the fact our adopted distances are calculated based on the empirical
cluster-based fits from Beers et al. (2000), taking into account
spectroscopic measurements of metallicity and surface gravity in order to
assign luminosity classes, while those on the right panel come from
application of a simple polynomial, which naturally leads to lack of
scatter.

The lower left panel of Fig.~\ref{fig:mag_comp_S10} shows the difference in
the assigned $M_r$ absolute magnitudes that arises when one compares the
revised C10 estimates with those of S10 for stars spectroscopically
classified as D stars ($\log g \geq 4.0$). For stars with [Fe/H] $> -2.0$,
the revised C10 determinations are brighter by a median offset of 0.38 mags
(rms 0.19 mags) for $0.4 < g-i < 0.8$, while the median offset of revised
C10 absolute magnitudes is 0.45 mags (rms 0.20 mags) brighter for bluer
stars in the range $g-i < 0.4$. The offsets are even larger for stars with
[Fe/H] $ < -2.0$. For the redder stars with $0.4 < g-i < 0.8$, the median
offset of the revised C10 determinations compared with S10 is 0.45 mags
(rms 0.16 mags) brighter; for bluer stars with $g-i < 0.4$, the median
offset is 0.52 mags (rms 0.18 mags) brighter.

The lower right panel of this figure shows the fractional difference in the
derived distances between the revised C10 and S10 scales. For stars with
[Fe/H] $> -2.0$ and $0.4 < g-i < 0.8$, the median offset of the revised C10
distances with respect to the S10 distances is 19\% (rms 10\%). In the
bluer range, $g-i < 0.4$, the offset increases to about 23\% (rms 11\%).
For stars with [Fe/H] $< -2.0$ and $0.4 < g-i < 0.8$, the median offset of
the revised C10 distances with respect to the S10 distances is 23\% (rms
10\%). In the bluer range, $g-i < 0.4$, the offset is 27\% (rms 11\%). All
distance differences are in the sense that the revised C10 scale is (as
expected) longer than the S10 scale.

\section{Absolute Magnitudes and Distances Based on Alternative Schemes}

Since much of the discord between the conclusions reached by C10 and S10
arise from their adopted absolute magnitudes and distances, we now consider
two additional approaches for obtaining estimates of these quantities. It
is worth keeping in mind that these comparisons are only valid for stars
that are confidently assigned D status, for which we enforce the
requirement that they have spectroscopic gravity estimates assigned by the
SSPP of $\log g \ge 4.0$.

\subsection{The Empirical Calibration of I08}

We first consider the relationship adopted by I08, as summarized by their
Eqn. (A7), used in conjunction with the metallicity correction in their
Eqn. (A2) and Eqn. (A3). When combined into a single equation, one obtains:

\begin{eqnarray}
M_{r}(g-i, [\rm Fe/H]) &=& -0.56 + 14.32\,x -12.97\,x^{2} \nonumber \\
                       && + 6.127\,x^{3}-1.267\,x^{4}+0.0967\,x^{5} \nonumber \\
                       && -1.11\,[\rm Fe/H]-0.18\,[\rm Fe/H]^{2}, 
\end{eqnarray}
\noindent where $x=(g-i)$.  This was the recommended final photometric
parallax relationship from I08, where it is claimed to be valid (for
main-sequence stars) over a wide color range ($0.2 < g-i < 4.0$).

The S10 study did not make use of the above equation, but rather, adopted an absolute
magnitude relationship taken from a previous stage of the I08 analysis, given there as
Eqn. (A1), and applied a metallicity correction from Eqn.(A2) and Eqn. (A3) to obtain:

\begin{eqnarray}
M_r(g-i, [\rm Fe/H])  &=& 1.65 + 6.29\,x -2.30\,x^2 \nonumber \\
                      && -1.11\,[\rm Fe/H] -0.18\,[\rm Fe/H]^2, 
\end{eqnarray}
\noindent where $x=(g-i)$.

The S10 paper argued that their adopted absolute magnitude determinations
agreed better with their preferred set of isochrones (the BaSTI isochrones:
Pietrinferni et al. 2004, 2006), but in fact I08 did not expect this
relationship (which is from an early step in their development of the
appropriate absolute magnitude prediction) to perform well for bluer stars
near the main-sequence turnoff. This is a crucial limitation, as the
calibration-star sample considered by C07 and C10 includes a considerable
number of bluer objects -- 19\% of the C10 sample, for example, have $g-i <
0.4$. The fraction becomes even larger at low metallicity -- 31\% for
[Fe/H]$ < -1.0$, and 46\% for [Fe/H]$ < -2.0$. This relationship also does
not take into account corrections for differing ages of the underlying
stellar populations that were applied by I08 in seeking a more generally
useful photometric parallax method. The combination of these two effects
accounts for much of the discrepancy cited by S10 in the absolute
magnitudes (hence distances) used by the C07 and C10 studies.

\begin{figure*}
\centering
\includegraphics[angle=90,scale=0.57]{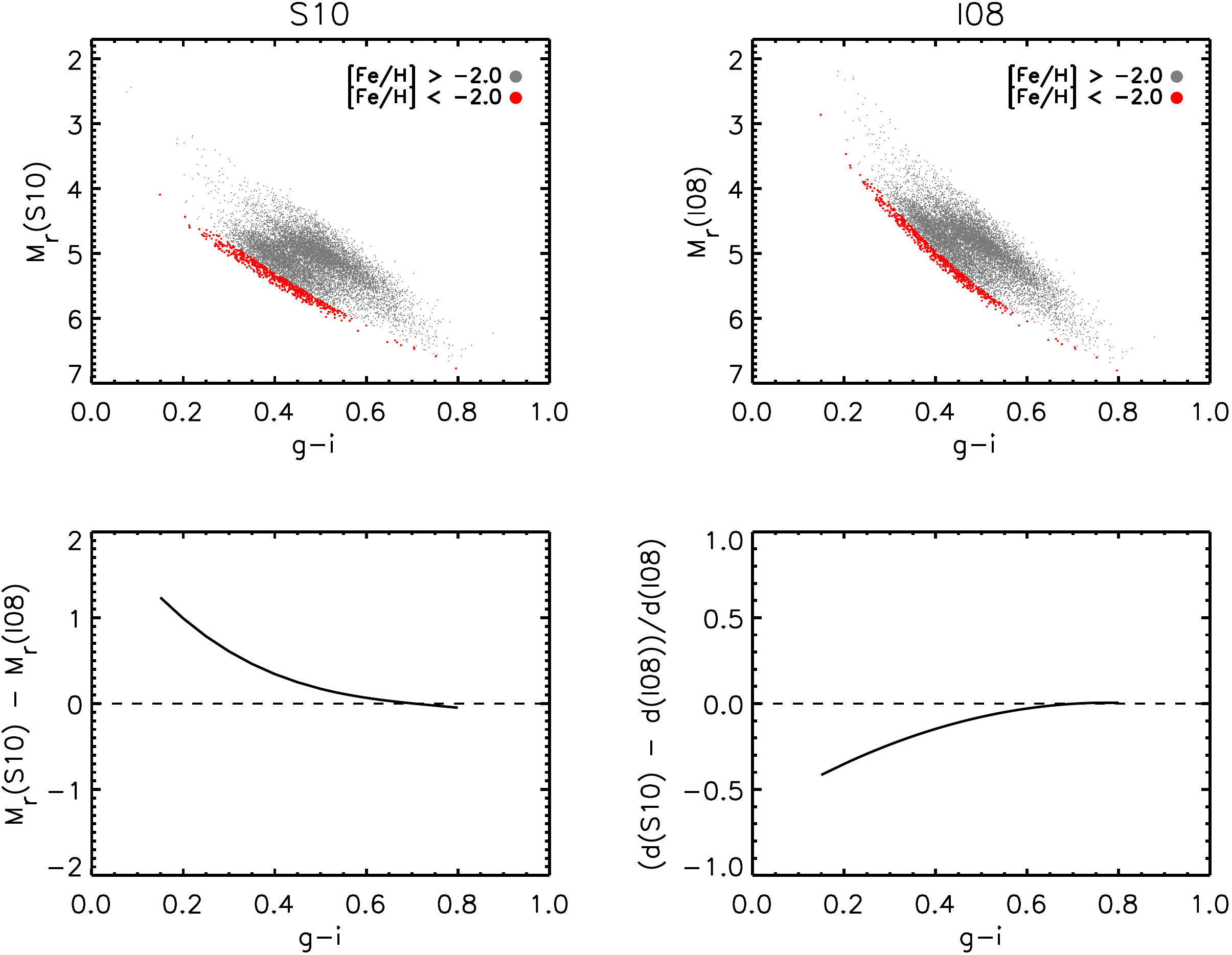}
\caption{{\it Upper left}: M$_r, g-i$ CMD for stars with
spectroscopically assigned D classifications, with absolute magnitudes
calculated from Eqn. (A1) of I08, as adopted by S10. {\it Upper right}: M$_r,
g-i$ CMD for stars with spectroscopically assigned D classifications, with
absolute magnitudes calculated from Eqn. (A7) of I08, as adopted by I08. The
stars with [Fe/H] $> -2.0$ are shown as gray dots, while those with [Fe/H] $<
-2.0$ are shown as red dots. {\it Lower left}: Difference between the M$_r$
absolute magnitudes for stars with spectroscopically assigned D classifications
for the S10 and I08 calculations, as a function of $g-i$. {\it Lower right}:
Fractional change in derived distances from those adopted by S10 as compared to
those adopted by I08, as a function of $g-i$.}
\label{fig:mag_comp_I08}
\end{figure*}

The upper left panel of Fig.~\ref{fig:mag_comp_I08} shows the CMD for stars
with spectroscopic assignments of D ($\log g \geq 4.0$), with absolute
magnitudes assigned by the relationship adopted by S10 (Eqn. 3 above). The
upper right panel shows the corresponding CMD obtained using the absolute
magnitudes from Eqn. 2 above, which is the recommended relationship from
I08. Note that in the evaluation of both relationships above, the
[Fe/H]$_C$ from C10 was employed, although similar results are obtained
when either the photometric metallicity estimates from I08 or the adopted
metallicity from the SSPP ([Fe/H]$_A$) are used. The stars are color-coded
to indicate metallicities above and below [Fe/H] $ = -2.0$. Note, however,
that since both procedures adopted the same metallicity correction scheme,
there are {\it no differences} in their behavior over different intervals
of [Fe/H].

The lower left panel of Fig.~\ref{fig:mag_comp_I08} shows the difference in
the assigned $M_r$ absolute magnitudes that arises when one compares the
adopted S10 and I08 relationships. For stars with $0.4 < g-i < 0.8$, the
median offset is 0.23 mags, with the S10 assignments being fainter. The
difference for bluer stars with $g-i < 0.4$ range from $\sim 0.23$ mags
fainter at the red end of this interval to roughly 1.0 mags fainter at the
blue end (median difference of 0.48 mags).

The lower right panel of this figure shows the fractional difference in the
derived distances between S10 and I08. For redder stars with $0.4 < g-i <
0.8$, the difference amounts to no more than about 15\% at the blue end of
this range (median offset of 10\%), but for the bluer stars with $g-i <
0.4$ the difference increases from $\sim 15$\% up to roughly 40\%, with a
median offset of 20\%. All distance differences are in the sense that the
S10 scale is shorter than the I08 scale.

\subsection{The Calibrated Isochrone Approach}

Distances to individual stars can also be estimated using a set of stellar
isochrones, once they have been properly calibrated against the observed
colors and magnitudes of stars with known distances and ages. For the
present exercise, we follow the prescription in An et al. (2009b) to derive
distances to individual stars employing stellar isochrones with empirical
corrections on the colors (An et al. 2009a). This calibration was based on
photometry from An et al. (2008) for a number of open and globular
clusters, including M67 ([Fe/H] $ = 0.0$) and M92 ([Fe/H] $ = -2.4$), which
provides metallicity-dependent color corrections in $ugriz$ over the
metallicity range under consideration. A full description of the isochrone
calibration can be found in A11.

After correcting the photometry for dust extinction, we performed model
fits over the full parameter space (with metallicity range $-3.0 \leq$
[Fe/H] $\leq +0.4$). We included $griz$ photometry and the key SSPP
atmospheric parameters ([Fe/H], \logg\ , $T_{\rm eff}$) in the model fits,
and found a best-fitting model by searching for a minimum $\chi^2$ of the
fit. Note that, for consistency with the other approaches, the corrected
metallicity [Fe/H]$_C$ was employed. We assumed minimum errors in the
photometry of $0.01$~mags for $gri$ and $0.02$~mags for $z$, and took
conservative errors of $0.3$~dex for [Fe/H], $160$~K for $T_{\rm eff}$, and
$0.4$~dex for \logg, as characteristic errors in each of these parameters
(including possible systematic scale differences between the SSPP and the
models). The lower limit of [Fe/H] in the models is $-3.0$, so we assumed
[Fe/H] $ = -3.0$ for any stars with metallicity less than this value. This
choice has a negligible impact on distance estimation, since the isochrones
are insensitive to a change in the atmospheric abundances for [Fe/H] $ <
-3.0$. An age of $12$~Gyr is assumed for [Fe/H] $ < -1.0$, while $4$~Gyr is
taken for [Fe/H] $ > -0.3$, with a linearly interpolated value for
metallicities between the two boundaries. Solutions for distances were
dropped from further consideration in cases where either the fitting
process did not converge, or if the final reduced $\chi^2$ of a converged
fit exceeded 1.2.

Unlike the original approach described by An et al. (2009b), the calibrated
isochrones actually reach into the main-sequence turnoff region, thus
distance estimates are available for both TO and SG stars, in addition to D
stars, albeit with lower accuracy in the distance estimates. For the
purpose of our present comparisons we only accepted stars with
spectroscopic assignments of surface gravity $\log g \geq 4.0$. An
inter-comparison of results from various color indices indicates that the
internal error in the distance modulus is $\sim0.1$~mag; an additional
$\sim0.1$~mag error is expected from the errors in age, [Fe/H],
[$\alpha$/Fe], and adopted E($B - V$). This suggests that the associated
distance-modulus error is $\sim0.1-0.2$~mags for individual stars. As was
the case for the I08 approach, the effects of binarity are more difficult
to quantify, and are not included in this error estimate (see An et al.
2007).

\begin{figure*}
\hspace{-4.5cm}
\includegraphics[angle=90,scale=0.75]{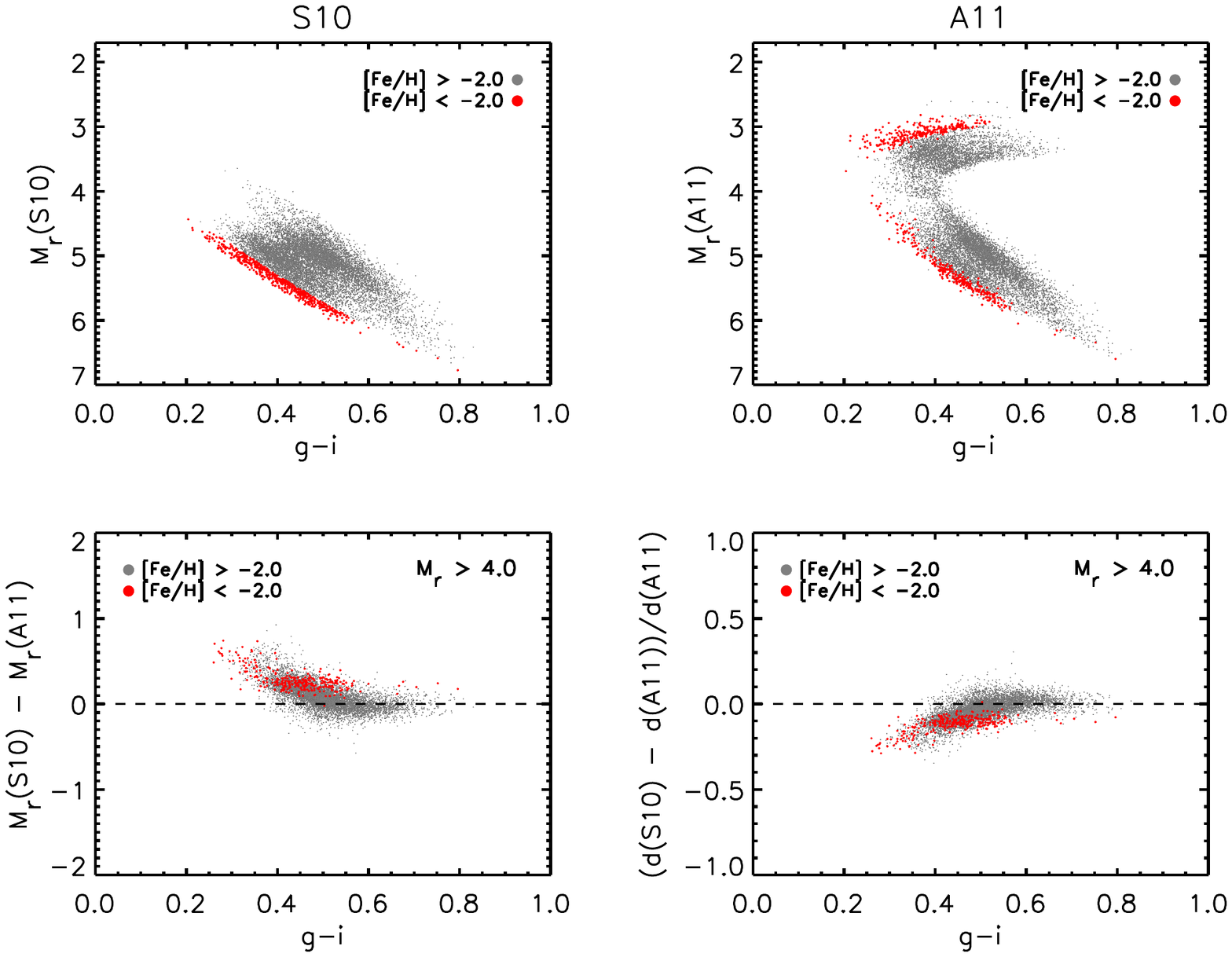}
\caption{{\it Upper left}: M$_r, g-i$ CMD for stars with
spectroscopically assigned D classifications, with absolute magnitudes
calculated from Eqn. (A1) of I08, as adopted by S10. {\it Upper right}: M$_r,
g-i$ CMD for stars with spectroscopically assigned D classifications, with
absolute magnitudes calculated using the calibrated isochrone fitting procedures
of A11. The stars with [Fe/H] $> -2.0$ are shown as gray dots, while those with
[Fe/H] $< -2.0$ are shown as red dots. {\it Lower left}: Difference between the
M$_r$ absolute magnitudes for stars with spectroscopically assigned D
classifications for the S10 and A11 calculations, as a function of $g-i$. {\it
Lower right}: Fractional change in derived distances from those adopted by S10
as compared to those adopted by A11, as a function of $g-i$.}
\label{fig:mag_comp_An}
\end{figure*}

The upper left panel of Fig.~\ref{fig:mag_comp_An} shows the CMD obtained
using the absolute magnitudes from Eqn. (A1) of I08 as adopted by S10.
The upper right panel shows the CMD for stars with spectroscopic
assignments as D ($\log g \geq 4.0$), with absolute magnitudes assigned by
the calibrated isochrone procedure of A11. Note that in the evaluation of
both relationships above, the [Fe/H]$_C$ used by C10 was employed, although
similar results are obtained when either the photometric metallicity
estimates or the adopted metallicity from the SSPP ([Fe/H]$_A$) were used.
The stars are color-coded to indicate metallicities above and below [Fe/H]$
= -2.0$. As is clear from inspection of the upper right panel, the A11
procedure assigns roughly half of the spectroscopic D stars into SG/G
classifications, with correspondingly brighter absolute magnitudes near
M$_r \sim 3$.

The lower left panel of Fig.~\ref{fig:mag_comp_An} shows the difference in
the assigned $M_r$ absolute magnitudes that arises when one compares the
adopted S10 and A11 relationships for stars spectroscopically classified as
D stars. For the purpose of this exercise we focus on the stars to which
the A11 procedure assigns dwarf status, with absolute magnitudes M$_r >
4.0$. For stars with [Fe/H] $> -2.0$, the S10 determinations are fainter
than those of A11 by a median offset of 0.10 mags (rms 0.09 mags) for $0.4
< g-i < 0.8$, while they are fainter by up to 0.7 mags (median offset of
0.31 mags, rms 0.15 mags) for the bluer stars with $g-i < 0.4$. The offsets
are significantly larger for stars with [Fe/H] $ < -2.0$. For the redder
stars with $0.4 < g-i < 0.8$, the median offset of the S10 determinations
compared with A11 is 0.24 mags (rms 0.06 mags) fainter; for bluer stars
with $g-i < 0.4$, the median offset is 0.41 mags (rms 0.15 mags) fainter.

The lower right panel of this figure shows the fractional difference in the
derived distances between S10 and A11 scales. For stars with [Fe/H] $>
-2.0$ and $0.4 < g-i < 0.8$, the median offset of the S10 distances with
respect to the A11 distances is only about 4\% (rms 4\%). In the bluer
range, $g-i < 0.4$, the median offset increases to about 13\% (rms 6\%).
For stars with [Fe/H] $< -2.0$ and $0.4 < g-i < 0.8$, the median offset of
the S10 distances with respect to the A11 distances increases to 10\% (rms
3\%). In the bluer range, $g-i < 0.4$, the median offset increases to about
17\% (rms 6\%). All distance differences are in the sense that the S10
scale is shorter than the A11 scale.

\subsection{Comparison with the C10 Dwarfs}

\begin{figure*}
\hspace{-4.5cm}
\includegraphics[angle=90,scale=0.75]{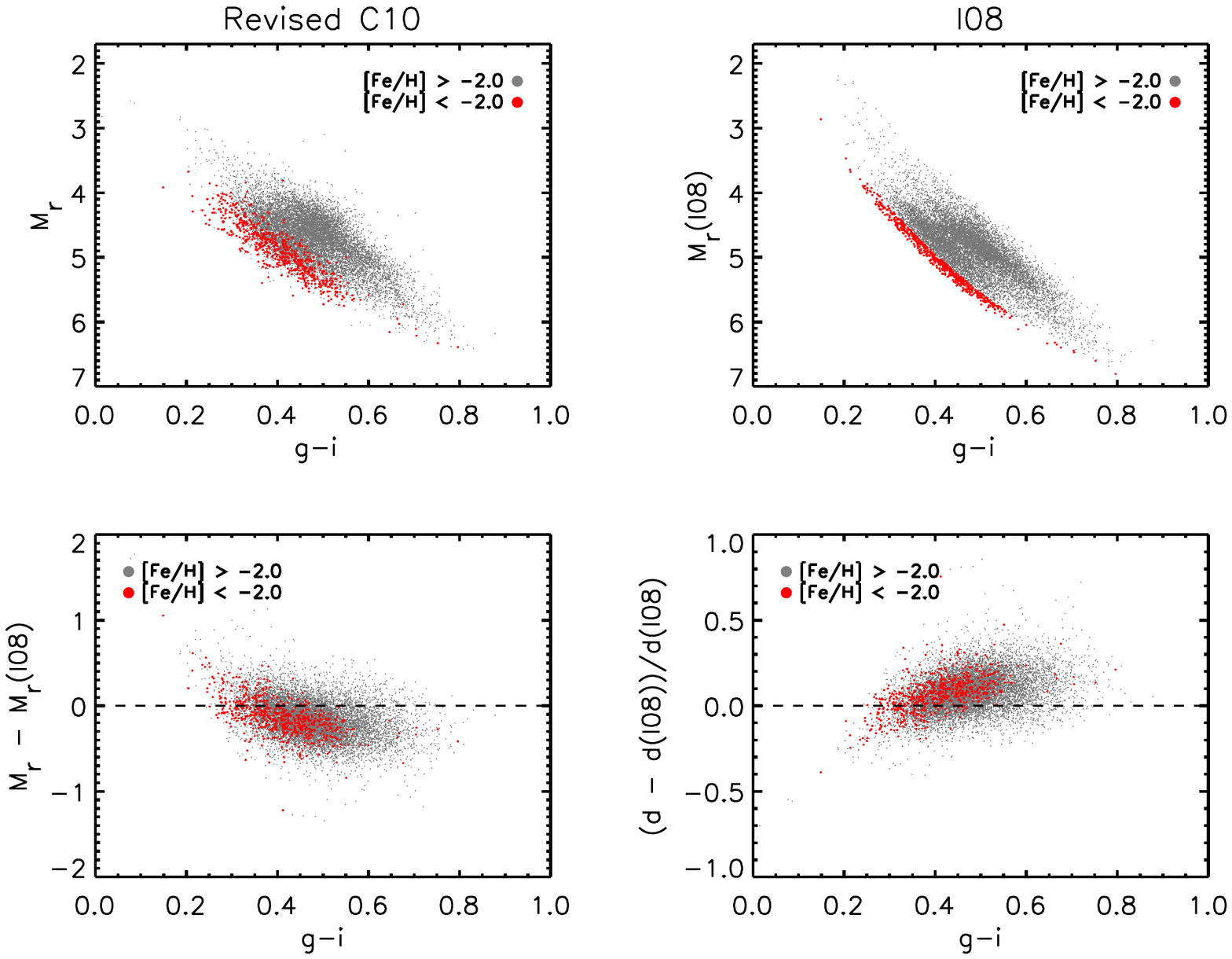}
\caption{{\it Upper left}: M$_r, g-i$ CMD for stars with revised C10
luminosity classifications and with spectroscopically assigned D
classifications, as a function of $g-i$. {\it Upper right}: M$_r, g-i$ CMD for
stars with spectroscopically assigned D classifications, with absolute
magnitudes calculated from Eqn. (A7) of I08, as adopted by I08. The stars with
[Fe/H] $> -2.0$ are shown as gray dots, while those with [Fe/H] $< -2.0$ are
shown as red dots. {\it Lower left}: Difference between the M$_r$ absolute
magnitudes for stars with spectroscopically assigned D classifications for the
A11 and I08 calculations, as a function of $g-i$. {\it Lower right}:
Fractional change in the revised distances from C10 as compared to those adopted
by I08, as a function of $g-i$.}
\label{fig:mag_comp_C10_I08}
\end{figure*}

\begin{figure*}
\hspace{-4.5cm}
\includegraphics[angle=90,scale=0.75]{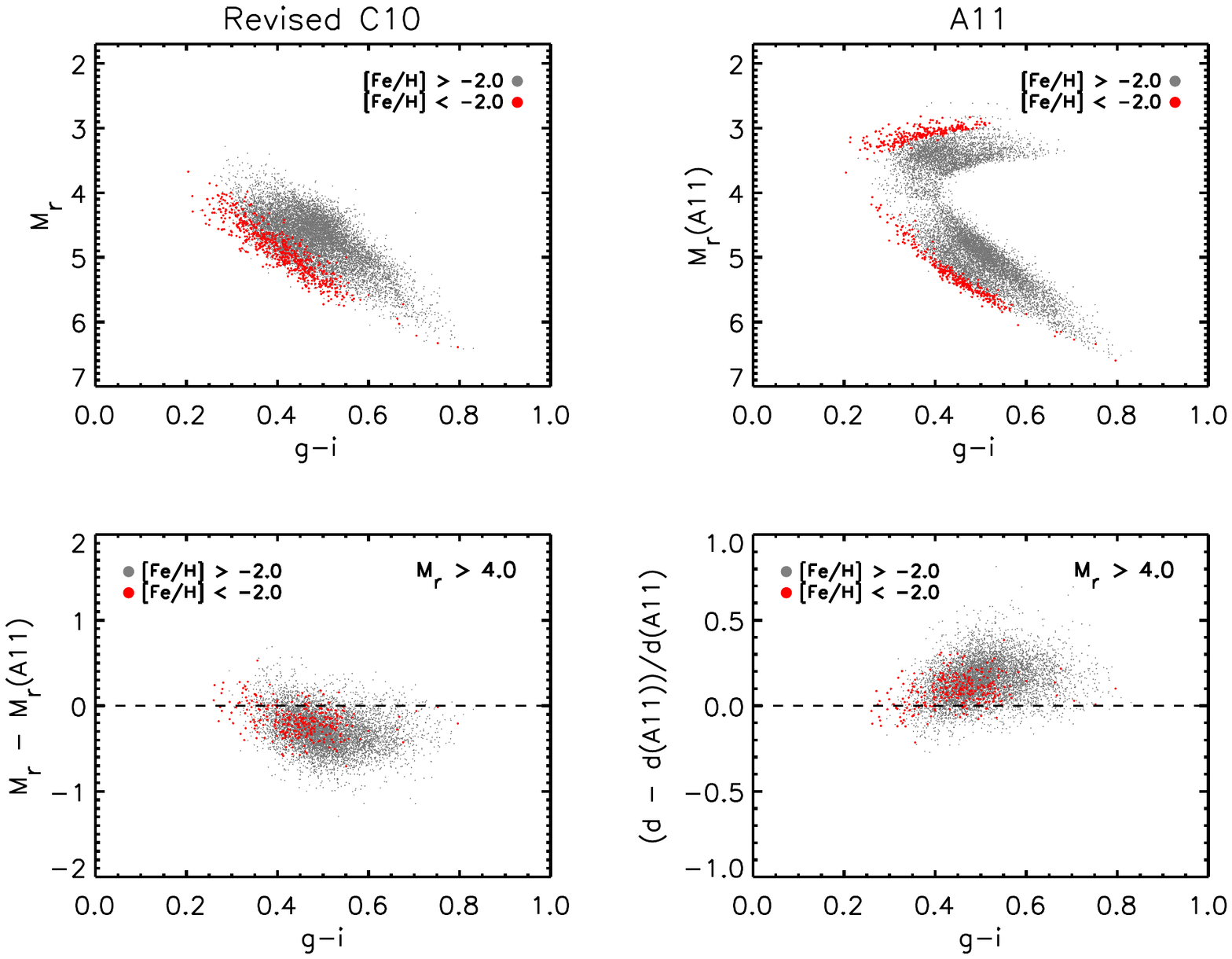}
\caption{{\it Upper left}: M$_r, g-i$ CMD for stars with revised C10
luminosity classifications and with spectroscopically assigned D
classifications, as a function of $g-i$. {\it Upper right}: M$_r, g-i$ CMD for
stars with spectroscopically assigned D classifications, with absolute
magnitudes calculated from Eqn. (A7) of I08, as adopted by I08.
The stars with [Fe/H] $> -2.0$ are shown as gray dots, while those with [Fe/H]
$< -2.0$ are shown as red dots. {\it Lower left}: Difference between the M$_r$
absolute magnitudes for stars with spectroscopically assigned D classifications
for the revised C10 and A11 calculations, as a function of $g-i$. {\it Lower
right}: Fractional change in the revised distances from C10 as compared to those
adopted by A11, as a function of $g-i$.}
\label{fig:mag_comp_C10_An}
\end{figure*}

We now compare the C10 sample, with revised TO classifications, with the
calculations of I08 (Fig.~\ref{fig:mag_comp_C10_I08}) and with those of A11
(Fig.~\ref{fig:mag_comp_C10_An}). As can be appreciated by inspection of
these figures, the absolute magnitude scale for the revised C10 sample
agrees well with those from both I08 and A11 (in the latter case, one can
only consider the stars considered dwarfs by the A11 procedure; see below).

The lower left panel of Fig.~\ref{fig:mag_comp_C10_I08} shows the
difference in the assigned $M_r$ absolute magnitudes that arises when one
compares the revised C10 estimates with those of I08 for stars
spectroscopically classified as D stars. For stars with [Fe/H] $> -2.0$,
the revised C10 determinations are brighter by a median offset of 0.21 mags
(rms 0.16 mags) for $0.4 < g-i < 0.8$, while the median offset of revised
C10 absolute magnitudes is 0.14 mags (rms 0.27 mags) brighter for bluer
stars in the range $g-i < 0.4$. The offsets are of similar size for stars
with [Fe/H] $ < -2.0$. For the redder stars with $0.4 < g-i < 0.8$, the
median offset of the revised C10 determinations compared with I08 is 0.23
mags (rms 0.15 mags) brighter; for bluer stars, the median offset is 0.13
mags (rms 0.14 mags) brighter.

The lower right panel of this figure shows the fractional difference in the
derived distances between the revised C10 and I08 scales. For stars with
[Fe/H] $> -2.0$ and $0.4 < g-i < 0.8$, the median offset of the revised C10
distances with respect to the I08 distances is 10\% (rms 9\%). In the bluer
range, $g-i < 0.4$, the median offset is about 6\% (rms 7\%). For stars
with [Fe/H] $< -2.0$ and $0.4 < g-i < 0.8$, the median offset of the
revised C10 distances with respect to the I08 distances is 11\% (rms 8\%).
In the bluer range, $g-i < 0.4$, the median offset is 6\% (rms 6\%). All
distance differences are in the sense that the revised C10 scale is longer
than the I08 scale.

Turning to Fig.~\ref{fig:mag_comp_C10_An}, if we focus on the stars that
are assigned dwarf status by the A11 procedure (we accomplish this by only
comparing stars with derived $M_r > 4.0$), the agreement between the
revised C10 estimates of absolute magnitude and distance is only slightly
worse, with respect to A11, than with respect to I08.

The lower left panel of Fig.~\ref{fig:mag_comp_C10_An} shows the difference
in the assigned $M_r$ absolute magnitudes that arises when one compares the
revised C10 estimates with those of A11, for stars spectroscopically
classified as D. For stars with [Fe/H] $> -2.0$, the revised C10
determinations are brighter by a median offset of 0.31 mags (rms 0.18 mags)
for $0.4 < g-i < 0.8$, while the median offset of revised C10 absolute
magnitudes is 0.17 mags (rms 0.15 mags) brighter for bluer stars in the
range $g-i < 0.4$. The offsets are smaller for stars with [Fe/H] $ < -2.0$.
For the redder stars with $0.4 < g-i < 0.8$, the median offset of the
revised C10 determinations compared with I08 is 0.21 mags (rms 0.14 mags)
brighter; for bluer stars with $g-i < 0.4$, the median offset is 0.15 mags
(rms 0.12 mags) brighter.

The lower right panel of this figure shows the fractional difference in the
derived distances between the revised C10 and A11 scales. For stars with
[Fe/H] $> -2.0$ and $0.4 < g-i < 0.8$, the median offset of the revised C10
distances with respect to the A11 distances is 15\% (rms 10\%). In the
bluer range, $g-i < 0.4$, the median offset decreases to about 8\% (rms
8\%). For stars with [Fe/H] $< -2.0$ and $0.4 < g-i < 0.8$, the median
offset of the revised C10 distances with respect to the I08 distances is
10\% (rms 7\%). In the bluer range, $g-i < 0.4$, the median offset is 7\%
(rms 6\%). All distance differences are in the sense that the revised C10
scale is longer than the A11 scale.

\subsection{Comparison Between A11 and I08}

\begin{figure*}
\hspace{-4.5cm}
\includegraphics[angle=90,scale=0.75]{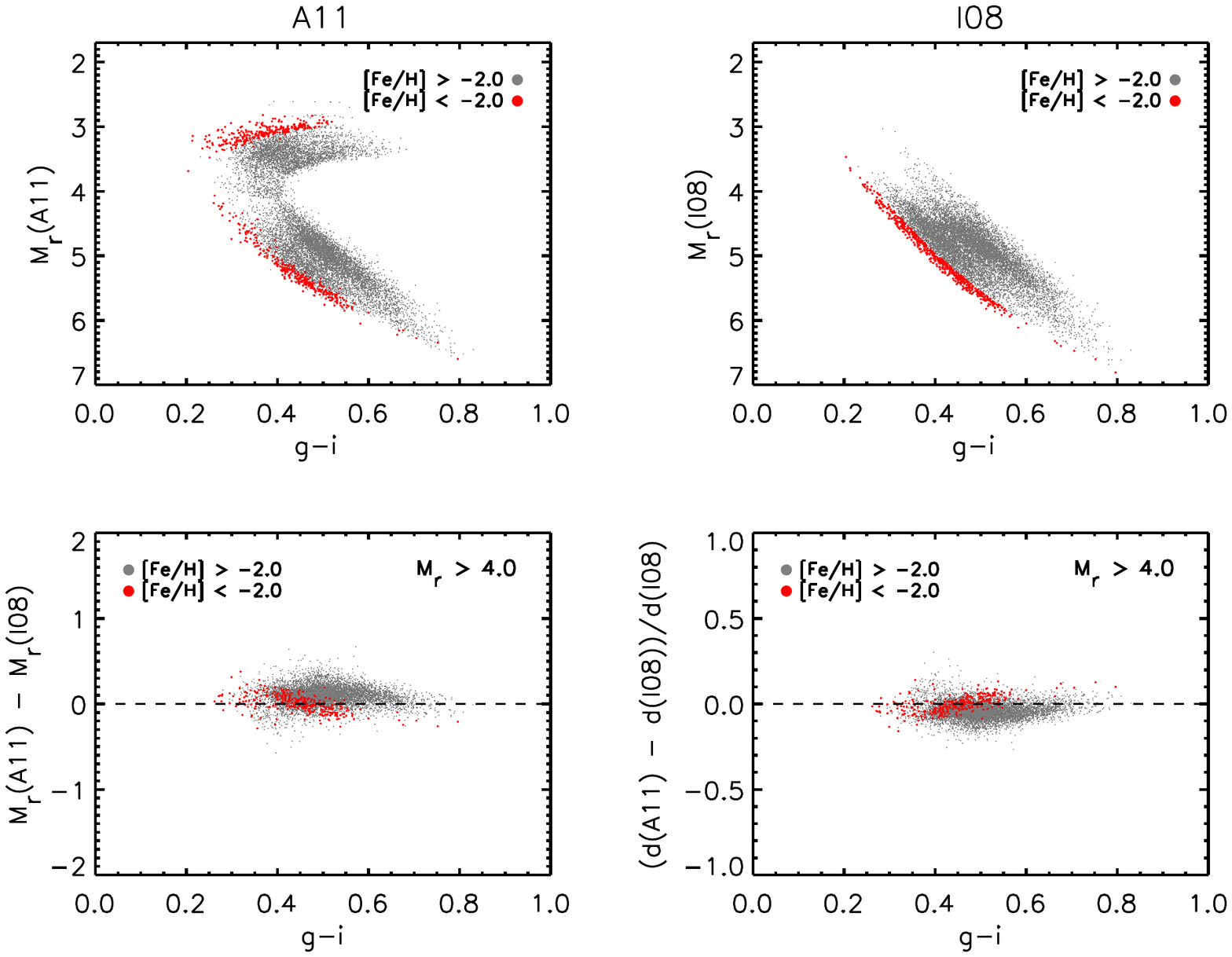}
\caption{{\it Upper left}: M$_r, g-i$ CMD for stars with
for stars with spectroscopically assigned D classifications, with absolute
magnitudes calculated using the calibrated isochrone fitting procedures of A11.
{\it Upper right}: M$_r, g-i$ CMD for stars with spectroscopically assigned D
classifications, with absolute magnitudes calculated from Eqn. (A7) of I08, as
adopted by I08. The stars with [Fe/H] $> -2.0$ are shown as gray dots, while
those with [Fe/H] $< -2.0$ are shown as red dots. {\it Lower left}: Difference
between the M$_r$ absolute magnitudes for stars with spectroscopically assigned
D classifications for the A11 and I08 calculations, as a function of $g-i$. {\it
Lower right}: Fractional change in derived distances from those adopted by A11
as compared to those adopted by I08, as a function of $g-i$.}
\label{fig:mag_comp_An_I08}
\end{figure*}

For completeness, Fig.~\ref{fig:mag_comp_An_I08} shows the comparison
between the isocohrone fitting procedure of A11 and the calculations of
I08.

The lower left panel of Fig.~\ref{fig:mag_comp_An_I08} shows the difference
in the assigned $M_r$ absolute magnitudes between the A11 and I08
estimates, for stars spectroscopically classified as D (and with $M_r >
4.0$, in order to only compare the stars considered as dwarfs by the A11
procedure). For stars with [Fe/H] $> -2.0$, the A11 determinations are
fainter by a median offset of 0.10 mags (rms 0.08 mags) for $0.4 < g-i <
0.8$, while the median offset is 0.12 mags (rms 0.08 mags) fainter for
bluer stars in the range $g-i < 0.4$. The offsets are smaller for stars
with [Fe/H] $ < -2.0$. For the redder stars with $0.4 < g-i < 0.8$, the
median offset of the A11 determinations compared with I08 is 0.06 mags (rms
0.06 mags) brighter; for bluer stars with $g-i < 0.4$, the median offset is
0.10 mags (rms 0.08 mags) fainter.

The lower right panel of this figure shows the fractional difference in the
derived distances between the A11 and I08 calculations. For stars with
[Fe/H] $> -2.0$ and $0.4 < g-i < 0.8$, the median offset of the A11
distances with respect to the I08 distances is 5\% (rms 4\%). In the bluer
range, $g-i < 0.4$, the offset is also about 5\% (rms 4\%). For stars with
[Fe/H] $< -2.0$ and $0.4 < g-i < 0.8$, the median offset of the A11
distances with respect to the I08 distances is 3\% (rms 3\%). In the bluer
range, $g-i < 0.4$, the offset is similar, about 4\% (rms 4\%). The
distance differences are in the sense that, for the redder stars, the A11
scale is longer than that of I08, while for the bluer stars, the A11 scale
is shorter than that of I08.

If we restrict our attention to the stars with [Fe/H] $< -2.0$, the ones
that matter the most for inferences concerning an outer-halo population, we
conclude from the above analysis that the I08 and A11 distance scales are
compatible with one another (maximum offsets of around 5\%), while the
revised C10 distance scale differs (in the sense of being longer) than both
the I08 and A11 scales by no more than about 10\% (better for stars near
the main-sequence turnoff, around 6-7\%). By contrast, the S10 scale
differs (in the sense of being shorter) with respect to the I08 scale by
between 10\% and 18\% (independent of metallicity; worse for stars near the
main-sequence turnoff), and similarly, between 10\% and 17\% (worse for
stars near the main-sequence turnoff) with respect to the A11 scale.
Although it is presently unknown which of these distance scales is closer
to ``ground truth'', the greater disagreement of the S10 scale (in
particular close to the main-sequence turnoff), not only with respect to
the revised C10 scale, but also with respect to those of I08 and A11,
suggests that it is the S10 scale that should be considered suspect, rather
than the revised C10 scale.

\section{A Reanalysis of Kinematics for Likely Outer-Halo Stars}

We now reconsider a limited kinematic analysis for a local sample of the
SDSS DR7 calibration stars following the procedures described by C10,
making use of the four different sets of distance assignments discussed
above for calculation of the full space motions. In order to provide a fair
comparison, we apply the same local volume constraints ($7 < R < 10$ kpc
and $d < 4$ kpc) to the various samples, but use the values of $R$ and $d$
that would be obtained for each of the different distance scales. This has
the obvious result that different numbers of stars will enter into each
sample. In order to maximize the contribution from proposed outer-halo
stars, we choose to only include stars with [Fe/H] $\le -2.0$. Our purpose
is to test the robustness of the retrograde signature that was criticized
by S10, which is most evident at low metallicity.

\begin{figure*}
\hspace{-0.5cm}
\includegraphics[angle=90,scale=0.75]{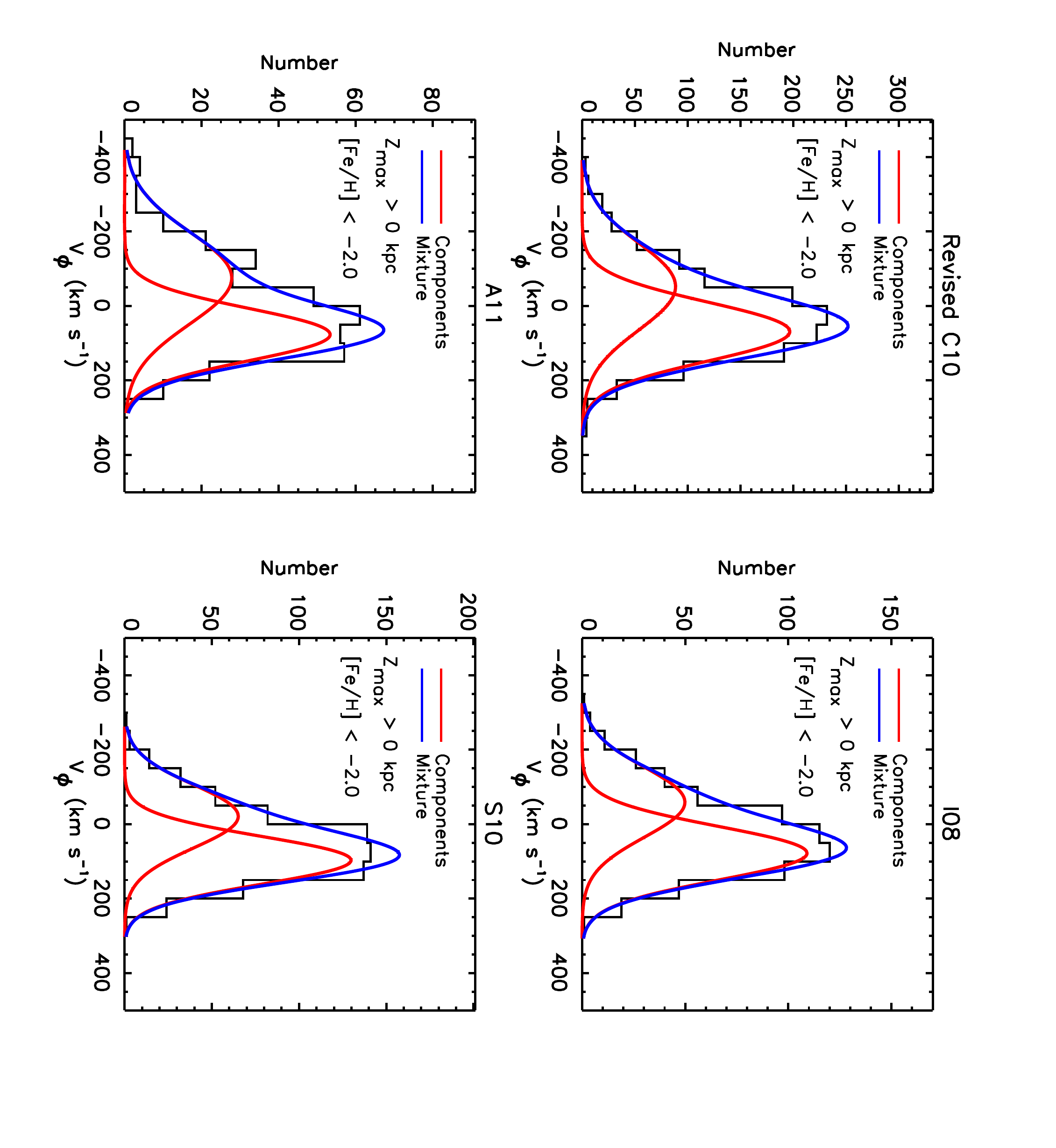}
\caption{{\it Upper left}: Histogram of \vphi\ for stars with revised C10
distances, with spectroscopically assigned D classifications, [Fe/H] $< -2.0$, and
all values of \zmax. The red solid lines are the suggested components from the
R-Mix procedure, while the blue solid line is the final mixture model.
{\it Upper right}: Similar, for D stars with I08 distances.
{\it Lower left}: Similar, for D stars with A11 distances.
{\it Lower right}: Similar, for D stars with S10 distances.}
\label{fig:v_phi_comp_zmax0}
\end{figure*}

Fig.~\ref{fig:v_phi_comp_zmax0} shows histograms of \vphi\ for the stars
spectroscopically classified as type D in the revised C10 sample, for all
ranges of \zmax\ (the maximum value of the distance above or below the
Galactic plane reached by a given star during its orbit). The red lines
shown in each panel are the two components of a model obtained by the R-Mix
procedure\footnote{{\tt http://www.math.mcmaster.ca/peter/mix/mix.html}} employed
by C10, to which the interested reader is referred for additional details.
As can be appreciated from inspection of this figure, all four of the
distance calibrations we consider lead to distributions of \vphi\ that
include asymmetric tails, which would not be expected to arise for a
single-component halo. Naturally, the suggested components and significance
of the splits vary from sample to sample; Table 2 summarizes these results.
Column (1) lists the sample under consideration (recall that the samples
differ only in their adopted distances as described above). Columns (2) and
(3) list the inferred means and dispersions (and their errors) of an
assumed Gaussian population for the first component of a two-component fit
to the observed distribution of \vphi, based on the R-Mix procedure.
Columns (4) and (5) list the same quantities for the second component
(where required). Column (6) is the p-value of the fits to a one-component
model.

The first section of Table 2 concerns the parameters of the R-Mix fits, for
D stars only, associated with Fig.~\ref{fig:v_phi_comp_zmax0}, which
applies to stars at all \zmax\ . Note that the number of dwarfs listed in
the revised C10 sample is more than twice that in the other samples; this
is the result of the inclusion of the reclassified TO $\rightarrow$ D
described above (including a subset of the stars with 3.75 $\le \log g <$
4.00). In the other samples, only the stars with spectroscopic estimates
$\log g \ge 4.0$ are included. From inspection of the table, the suggested
splits from R-Mix all include a retrograde and a prograde component, and
are highly statistically significant (in the sense that a one-component fit
is strongly rejected). This even includes the S10 sample, although one can
see that the formal derived velocity for the first component is less
retrograde than found for the other samples.  

\begin{figure*}
\hspace{-0.5cm}
\includegraphics[angle=90,scale=0.75]{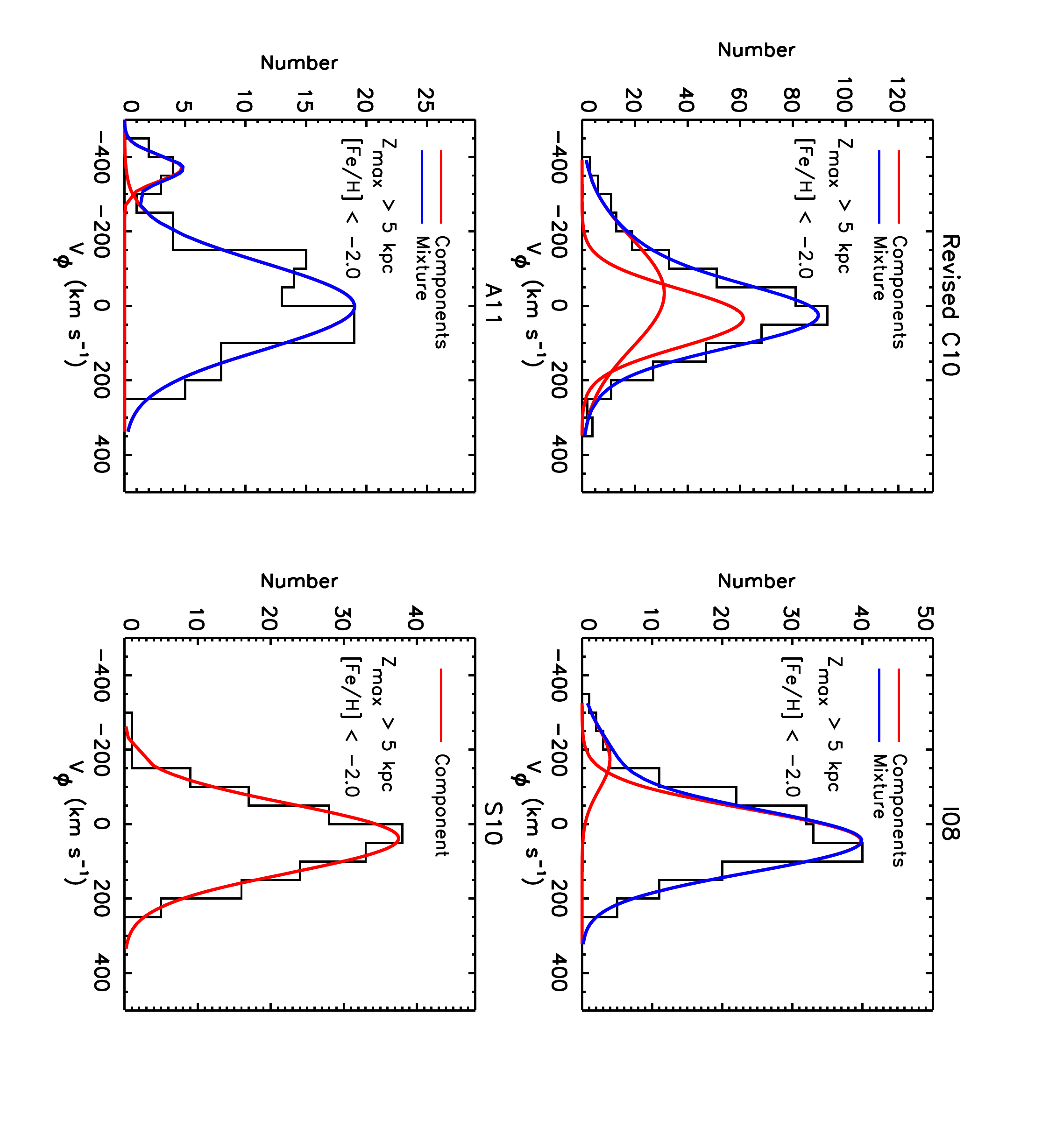}
\caption{{\it Upper left}: Histogram of \vphi\ for stars with revised C10
distances, with spectroscopically assigned D classifications, [Fe/H] $< -2.0$, and
\zmax\ $> 5$ kpc. The red solid lines are the suggested components from the
R-Mix procedure, while the blue solid line is the final mixture model.
{\it Upper right}: Similar, for D stars with I08 distances.
{\it Lower left}: Similar, for D stars with A11 distances.
{\it Lower right}: Similar, for D stars with S10 distances.}
\label{fig:v_phi_comp_zmax5}
\end{figure*}

Fig.~\ref{fig:v_phi_comp_zmax5} shows the result of a similar analysis for
the four different sets of distance calibrations, but restricted to only
include stars with derived estimates of \zmax\ $> 5$ kpc. The samples of
spectroscopically classified D stars on orbits that reach beyond 5 kpc from
the disk plane is much smaller than considered for all ranges of \zmax, but
the fraction of likely outer-halo stars included by this cut on \zmax\
should be increased.

Inspection of Fig.~\ref{fig:v_phi_comp_zmax5} reveals some interesting
differences. While the revised C10 sample (which is considerably larger
than the other samples) shown in the upper left panel exhibits a clear
asymmetric tail extending to negative \vphi, the tails of the I08 and A11
samples are weaker than previously, but located at larger negative values
of \vphi. We judge this to be primarily the result of the smaller numbers
of stars included. Of particular interest is the lower right panel, which
shows the result for the S10 sample. As can be seen, if one were to accept
the S10 absolute magnitude scale and corresponding distances, one would
indeed be driven to interpret at least this cut on the data as
well-represented by a single component, which was the essence of the
argument presented by S10.

The second section of Table 2 concerns the parameters of the R-Mix fits,
for D stars only, associated with Fig.~\ref{fig:v_phi_comp_zmax5}. From
inspection of the table, the suggested splits from R-Mix include a
retrograde and a prograde component for the revised C10 sample, the I08
sample, and the A11 sample, all of which are highly statistically
significant, but {\it not} for the S10 sample, which only allows for a
marginally prograde one-component fit. It is revealing that the inferred
prograde velocities for the second components have dropped considerably
from the case that considered all values of \zmax. Of course, it should be
kept in mind that the restriction here, for the purpose of comparison using
the D stars only, has resulted in rather small numbers of stars included
for the I08, A11, and S10 subsamples. For example, the split of the A11
sample to include a highly-retrograde, low-dispersion component is
presumably driven by small-number statistics.

\begin{figure*}
\hspace{-0.5cm}
\includegraphics[angle=90,scale=0.60]{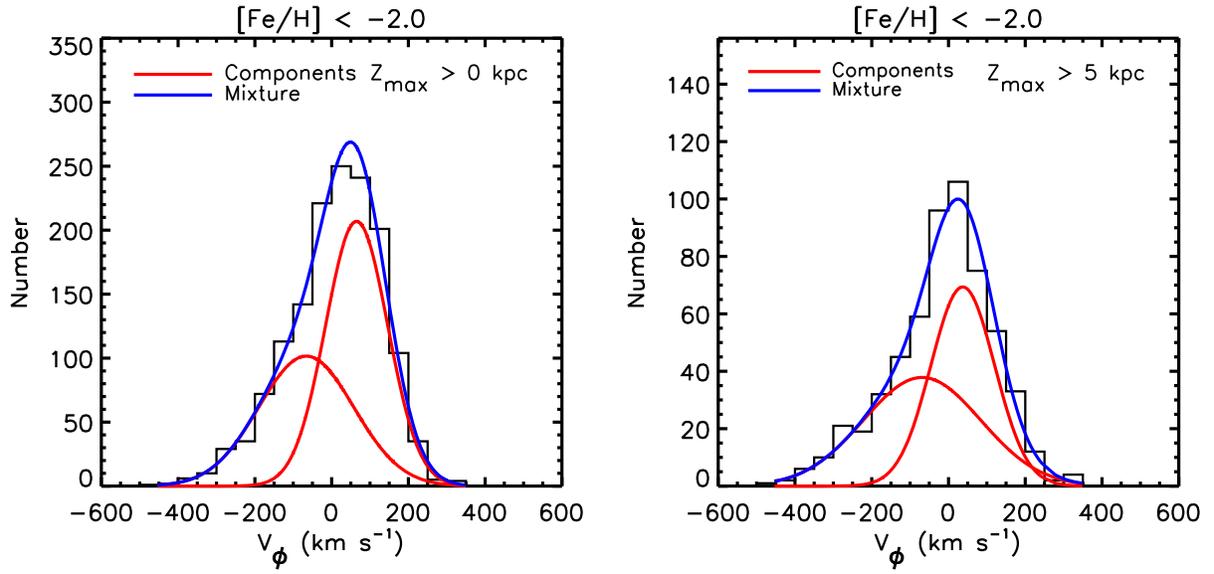}
\caption{{\it Left panel}: Histogram of \vphi\ for stars with revised C10
distances, with spectroscopically assigned D, TO, and SG/G classifications,
[Fe/H] $< -2.0$, and all values of \zmax. The red solid lines are the suggested
components from the R-Mix procedure, while the blue solid line is the final
mixture model. {\it Right Panel}: Similar, but for stars with \zmax\ $> 5$ kpc.}
\label{fig:v_phi_rmix}
\end{figure*}

Finally, we consider a similar set of analyses for the full revised C10
sample, including the D, TO, and SG/G classifications and their associated
distances and derived space motions. Fig.~\ref{fig:v_phi_rmix} shows the
results of this exercise for both the full range of \zmax\ (left panel) and
the case where only stars with \zmax\ $> 5$ kpc are considered. Inspection
reveals the clear presence of an asymmetric tail towards negative \vphi\ in
both cases, which we associate with the outer-halo component, as also
concluded by C07 and C10.

The last two sections of Table 2 apply to the samples shown in
Fig.~\ref{fig:v_phi_rmix}. As can be seen from inspection of this table,
the mean velocity of the retrograde component is similar to that obtained
by C10 for \zmax\ $> 5$ kpc, albeit with a slightly larger formal error
($-94 \pm 23$ \kms\ vs. $-80 \pm 13$ \kms). The dispersions of the
components are also similar to those obtained previously. A one-component
halo is strongly rejected in both cases. In all of the above, it should be
recalled that the final results given by C10 for the parameters of the
various suggested populations were derived with a custom maximum-likelihood
procedure, not from the R-Mix procedure described above. Hence, small
differences are expected in the final derived values.

Finally, it is worth recalling that Deason et al. (2011) speculated
that the retrograde signature they find for a large sample of
low-metallicity SDSS BHB stars (see further discussion below) could
be due to an incorrect adopted value for the LSR rotation velocity.
However, from inspection of the lower portion of Table 2, one
notes that significant {\it differences} in the mean rotational velocities
appear, indicating that a velocity shear is present between the presumed
underlying populations, as it is for the Deason et al. (2011) sample as
well. Thus, regardless of whether one assigns physical meaning to the
presence of a truly retrograde sigature associated with the outer-halo
component, all indications suggest that there is indeed a difference
between the rotational properties of the inner-halo and outer-halo
components.

\section{Additional Tests for the Presence of a Kinematically and/or \\
Chemically Distinct Outer Halo}

The limited kinematic analysis carried out above is already strong evidence
for the need of more than a single-component halo for the Milky Way, and
provides insight as to why a dual-halo interpretation was not supported by
S10, when using their adopted absolute magnitude scale. Nevertheless,
additional tests of a complex halo model that are not strongly influenced
by the adopted distance scale (other than for sample selection) are useful
to carry out. In this section we consider four such pieces of evidence --
(1) The origin of the retrograde signature from the revised C10 D
classifications as well as for the full set of D, TO, and SG/G
classifications, (2) Changes in the as-observed MDF of the revised C10
sample (including stars without measured proper motions and located outside
the local samples considered in the kinematic analysis), (3) The observed
distribution of Galactocentric radial velocities for the well-selected
sample of Blue Horizontal-Branch (BHB) stars from SDSS DR8 discussed by Xue
et al. (2011), and (4) Changes in the as-observed MDF of the BHB sample
over different cuts in Galactocentric distance.

\subsection{Additional Evidence (1):\\  The Origin of the Retrograde Signature}

It is useful to ask if the single-halo hypothesis, e.g., a halo as
described by the best-fit kinematic model from Bond et al. (2010) (and
argued to be valid by S10) can be rejected even without making use of the
analysis of full space motions. The gist of the difficulty with the
single-halo hypothesis is the fact that the derived rotational velocity
distribution is asymmetric for stars with low [Fe/H] (this asymmetry is
already present for stars with [Fe/H] $< -1.5$, and becomes even stronger
for stars with [Fe/H] $< -2.0$). 

The fraction of low-metallicity stars with highly-retrograde motions
(V$_{\phi}$ $< -$200 \kms) in the SDSS/SEGUE DR7 calibration-star sample is
significantly larger than for those with highly-prograde motions. For stars with
[Fe/H] $< -$1.5 (and exploring Z$_{max} >$ 0 kpc), the fraction of stars
with highly-retrograde motions is 9\%, compared with 4\% of stars with
highly-prograde motions (V$_{\phi}$ $> $ 200 \kms). For stars with [Fe/H] $<
-$2.0, the fractions are 13\% highly retrograde compared with 5\% highly
prograde. For orbits reaching to larger distances from the Galactic plane,
Z$_{max} >$ 5 kpc, the asymmetry is even stronger (as expected), 16\%
compared with 5\% for [Fe/H] $< -$1.5, and 20\% compared with 6\% at [Fe/H]
$< -$2.0. This asymmetric behavior is present even when only
spectroscopically classified dwarfs are considered
(Fig.~\ref{fig:v_phi_comp_zmax0}, Fig.~\ref{fig:v_phi_comp_zmax5}), which
alleviates concerns about potential systematic distance errors associated
with the other stellar classifications.
 
Belief in the reality of the derived asymmetry in the rotation velocities
leads naturally to several important questions. For example, ``Are stars in
the highly-retrograde subsample different in any other measured property
than the rest of sample?,'' and ``Why do they possess such large inferred
retrograde velocities?.''

\begin{figure*}
\hspace{-3.5cm}
\includegraphics[angle=90,scale=0.70]{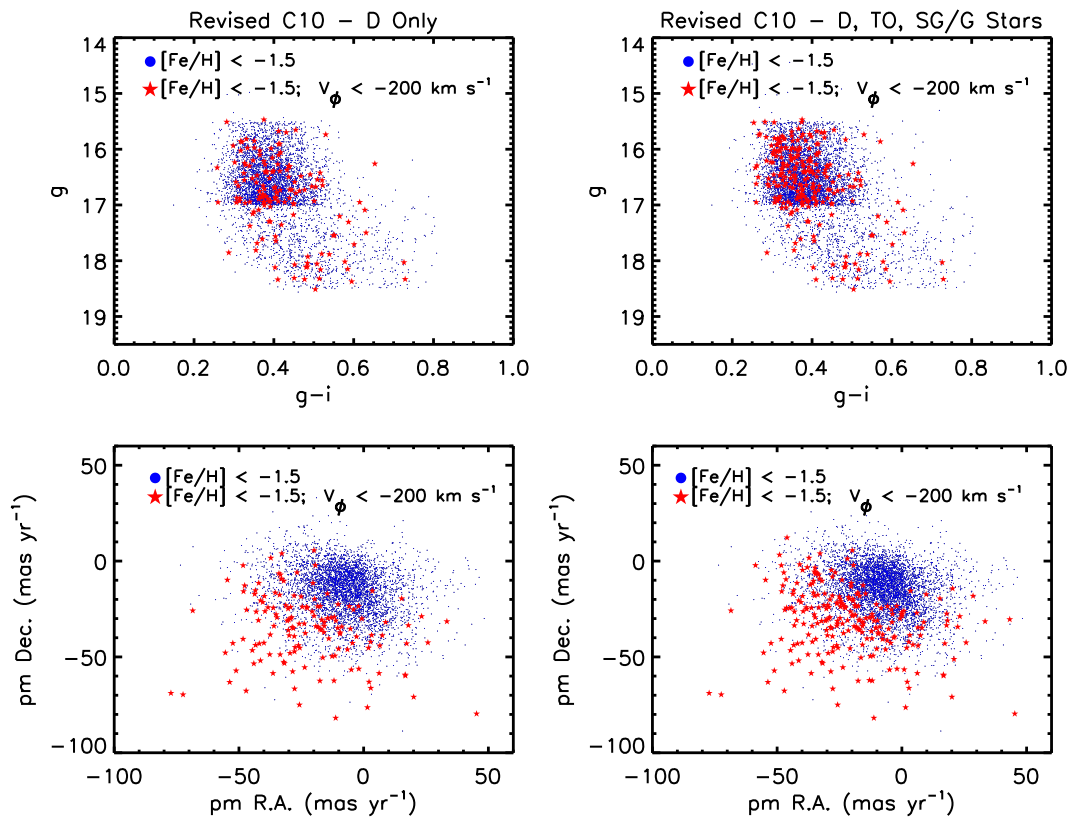}
\caption{{\it Upper left}: Apparent $g$-band magnitude vs. $g-i$ colors for the C10
stars with revised distances and [Fe/H] $< -1.5$, exclusively for stars
spectroscopically classified as D. The stars with highly-retrograde motions,
\vphi\ $< -200$ \kms, are indicated by the red stars; the rest of the sample 
is indicated with blue dots. The apparent structure as a function of $g$
magnitude in this diagram is due to the different selections used for the two classes of
calibration stars. {\it Upper right}: Similar, but for the full set of C10
classifications, D, TO, and SG/G. {\it Lower left}: The proper motion
distribution (vector components along the R.A. and Dec. directions) for the C10
stars with revised distances and [Fe/H] $< -1.5$, exclusively for stars
spectroscopically classified as D. The stars with highly-retrograde motions, \vphi\ $<
-200$ \kms, are indicated by the red stars; the rest of the sample is
indicated with blue dots. {\it Lower right}: Similar, but for the full set of
C10 classifications, D, TO, and SG/G.}
\label{fig:g_g-i_pm}
\end{figure*}

Fig.~\ref{fig:g_g-i_pm} shows that the distributions of the $g$-band apparent
magnitudes and $g-i$ colors are very similar for the full sample and the
highly-retrograde subsample (the large red squares highlight the subsample of
stars with highly-retrograde motion, \vphi\ $< -200$ \kms). Their distance
distributions are also similar (median distances of D stars are both $\sim$
2.1 kpc; median distances of the D, TO, and SG/G stars are both $\sim 2.5$
kpc). Since these are the quantities which, by and large, drive the
spectroscopic target selection, it is unlikely that spectroscopic selection
effects are important in this context. The apparent structure in this
figure (the discontinuity at $g = 17$) is simply the transition between the
two categories of calibration stars in the sample. The spectrophotometric
calibration stars cover the apparent magnitude range 15.5 $< g <$ 17.0, and
satisfy the color ranges 0.6 $< u-g <$ 1.2; 0.0 $< g-r <$ 0.6. The telluric
calibration stars cover the same color ranges as the spectrophotometric
calibration stars, but at fainter apparent magnitudes, in the range 17.0 $<
g <$ 18.5.

Although lower latitude stars (mostly arising from the SEGUE survey) are
present, the stars discussed here are observed at primarily high Galactic
latitudes (the median value of $|b|$ is $\sim 60^\circ$ for all subsamples
considered). Any presumed rotational signature has the greatest leverage at
lower latitudes. Hence, the concern that errors in the adopted distance
scale have ``amplified'' the derived rotational velocity component is
relieved somewhat by the distribution of the sample stars on the sky
themselves. Instead, the origin of the derived highly-retrograde motions is
primarily driven by their large (and asymmetric) measured proper motions
(bottom panels of Fig.~\ref{fig:g_g-i_pm}). The measured proper motions for
the highly-retrograde subsample are much larger than the random ($\sim 3-5$
mas yr$^{-1}$) and systematic ($<$ 1 mas yr$^{-1}$) proper motion errors.
These proper motion errors were determined using a sample of $\sim$60,000
quasars and are robust (see Section 2.3 of Bond et al. 2010). Of course,
this quasar-based analysis cannot exclude catastrophic errors (i.e., much
larger than expected from quasar behavior) in a small fraction of stars due
to effects such as a bad early-epoch plate, nearby bright stars with
diffraction spikes, large galaxies, etc. In order to minimize these
concerns, we have verified that the sky distribution of the 95 D stars (and
144 D, TO, and SG/G stars) in the highly-retrograde subsample, selected
with [Fe/H] $< -1.5$ and \vphi\ $< -200$ \kms, is similar to that for the
full sample (i.e., the highly-retrograde stars do not come from an isolated
small region, and issues such as chromatic differential aberration are
unlikely). In addition, we have visually inspected their SDSS images, and
found that essentially all are clean detections of isolated blue stars.
Based on this analysis, we conclude that there is no evidence that the
large observed proper motions for the highly-retrograde subsamples are due
to unrecognized systematic errors.

The visually apparent difference between the distributions of the
proper-motion components for the highly-retrograde subsample of stars and
the rest of the sample shown in Fig.~\ref{fig:g_g-i_pm} can be quantified
using the so-called Two Dimensional K-S test, described by Peacock (1983)
(see also Press et al. 1997). Application of the 2D K-S test clearly
rejects the null hypothesis that the stars identified as belonging to the
highly-retrograde tail in \vphi, on the basis of their derived space
motions, are drawn from the sample parent population as the rest of the
sample ($ p \ll 0.0001$). The discussion by S10 cautioned that asymmetries
in the derived azimuthal velocities might be artificially created due to a
Lutz \& Keller (1973) bias in the absolute magnitudes (hence distances) for a
given sample, which when applied to the proper motions could lead to the
presence of an extended counter-rotating tail in the distribution of \vphi.
While the possibility of such a bias exists, our tests based on the
observed proper motions alone greatly diminish the likelihood that the
highly-retrograde tail could have resulted from such an effect. That is, if
the retrograde signature were indeed created in the manner suggested by S10
alone (and the stars we assign to a highly-retrograde tail were otherwise
identical in their kinematic properties to the rest of the low-metallicity
halo stars), we would {\it a priori} expect their observed proper motions
to be drawn from the same parent population as those {\it not} in the tail.
This is clearly not what the data are telling us. 

In summary, the selection of stars by metallicity (a spectroscopic
quantity) generates a subsample with a derived asymmetric rotational
velocity distribution that is primarily due to the asymmetry of the
measured proper motions themselves (obtained from imaging data).
Although one can always raise the issue of selection effects in the SDSS
spectroscopic sample, any simple mechanism that would introduce the
observed behavior seems unlikely, because spectroscopic targeting of the
calibration-star sample is performed without direct knowledge of the proper
motion measurements. Therefore, this interplay between the independent
imaging and spectroscopic measurements is a strong argument that the
asymmetric \vphi\ distribution for low-metallicity stars is real, and that
it does not arise because of errors related to derived distances or other
effects we are aware of.

\subsection{Additional Evidence (2):\\ The Metallicity Distribution Function of
the C10 Sample with Revised Distances\\ and Variation with Distance from the
Galactic Plane}

The previous analyses of C07 and C10 both concluded that the MDF of the stars in
the SDSS calibration-star sample was inconsistent (for regions beyond the
possible influence of the disk system) with being drawn from a
location-invariant parent population, as would be demanded by the single-halo
hypothesis. Here we verify that this claim remains valid, even after
reassignment of a subset of the C10 TO stars into alternative luminosity
classifications. This is important to check because, as noted above, the
majority of these reassignments were TO $\rightarrow$ D, which clearly leads to a
reduction in their typical distances.

\begin{figure*}
\centering
\includegraphics[scale=0.75]{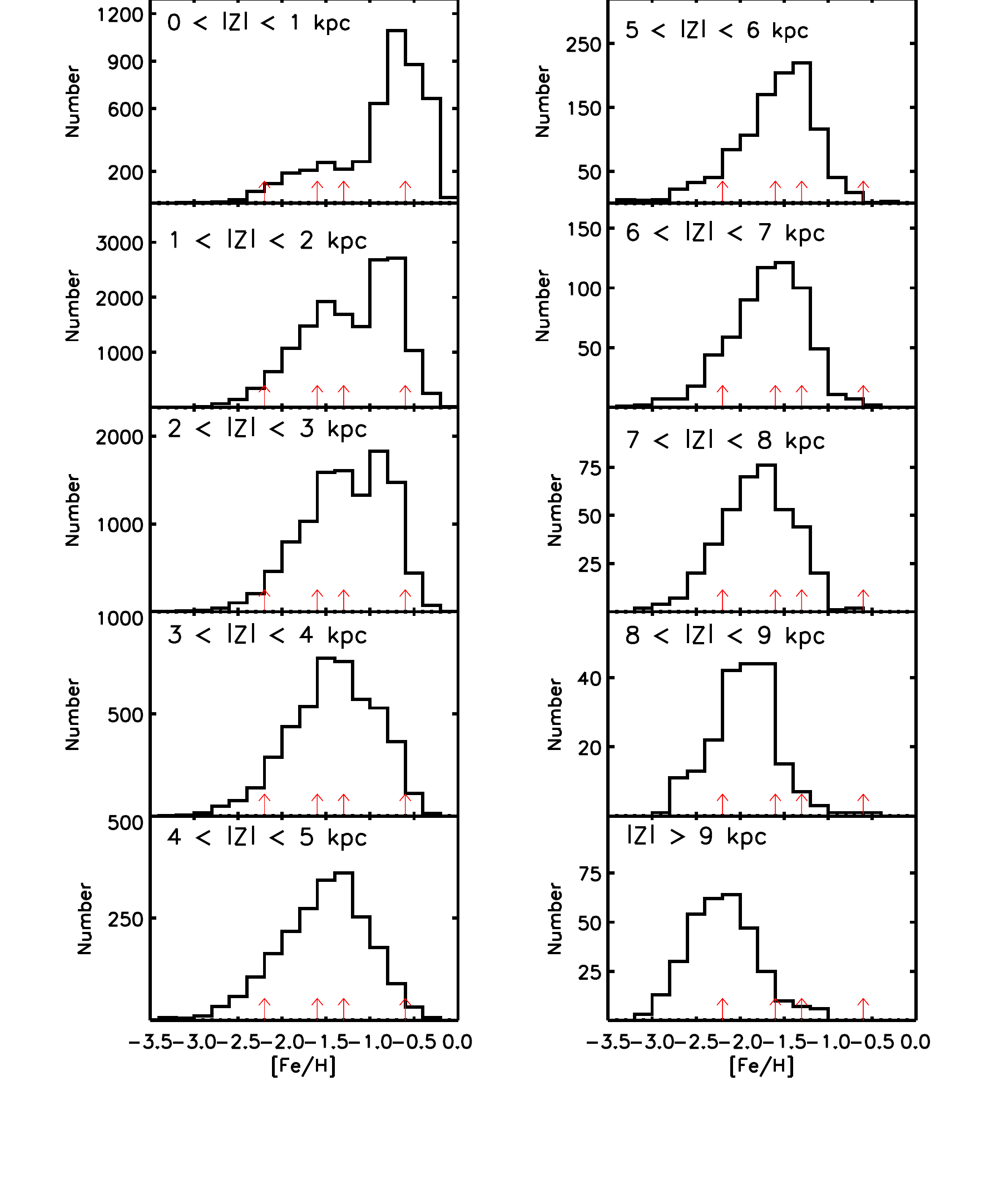}
\caption{As-observed metallicity distribution functions for stars from C10 with
revised distances, for various cuts in distance from the Galactic plane, $|Z|$.
The vertical red arrows mark the positions of the primary stellar components
modeled by C10, the thick disk ([Fe/H] $= -0.6$), the metal-weak thick disk
([Fe/H] $= -1.3$), the inner halo ([Fe/H] $= -1.6$), and the outer halo ([Fe/H]
$= -2.2$).}
\label{fig:mdf_zdist}
\end{figure*}

Fig.~\ref{fig:mdf_zdist} shows the as-observed MDF for the C10 sample with revised
distances, for cuts on distance from the Galactic plane, $|$Z$|$. This figure
exhibits strikingly similar behavior to that seen in, e.g., Fig.~20 of C10. The
SDSS/SEGUE DR7 calibration stars located within 5 kpc of the plane display MDFs
that are influenced primarily by the presence of the thick-disk, the metal-weak
thick disk, and the proposed inner-halo populations. In the ranges of distance
greater than 5 kpc, one sees a clear transition from the MDF of the proposed
inner-halo population, with peak metallicity near [Fe/H] $ = -1.6,$ to an MDF
dominated by progressively lower-metallicity stars, with a peak near [Fe/H] $ =
-2.2$, that are associated with the proposed outer-halo population.

From these first two additional pieces of evidence it is difficult to justify the single-halo hypothesis, either kinematically or chemically, unless other
attributes (such as smooth gradients of unknown physical origin in the motions
and metallicities of member stars in the SDSS/SEGUE calibration-star sample) are
invoked.

\subsection{Additional Evidence (3):\\ The Distribution of Galactocentric Radial
Velocities for BHB Stars from SDSS DR8 \\ and Variation with Metallicity}

The SDSS spectroscopic samples comprise a number of alternative
tracers that can be used to explore the nature of the Milky Way's halo system.
Among the most powerful are the BHB stars, which are intrinsically bright and
numerous, and have well-calibrated photometric distances. These have already
been used by a number of previous authors, including Yanny et al. (2000), Sirko
et al. (2004a,b), Xue et al. (2008, 2011), Bell et al. (2010), and Deason et al.
(2011), to explore various aspects of the nature of the Milky Way's stellar
halo.

The Xue et al. (2011) sample from the SDSS DR8 data release is of particular
value, because all of the constituent BHB stars have been classified based on
carefully applied spectroscopic tests of the Balmer lines, in addition to the
usual color cuts. It is a large (N $> 4000$ stars), well-controlled sample, with
available metallicities, radial velocities, and distance estimates, that samples
the inner and outer regions of the Galaxy at distances up to 80 kpc from the
Galactic center.

For the purpose of our present analysis we use the distance estimates for the
FHB stars reported by the SSPP, which in turn rely on the metallicity-dependent
calibration of the horizontal branch adopted by Beers et al. (2000). We employ
the metallicity estimate reported from the SSPP attributed to Wilhelm, Beers, \&
Gray (1999), which should be superior to alternative estimates for these warm
stars, and is accurate to on the order of 0.25-0.3 dex. The reported radial
velocities are expected to be accurate to better than 20 \kms, based on numerous
previous tests.

\begin{figure*}
\hspace{-0.5cm}
\includegraphics[scale=0.75]{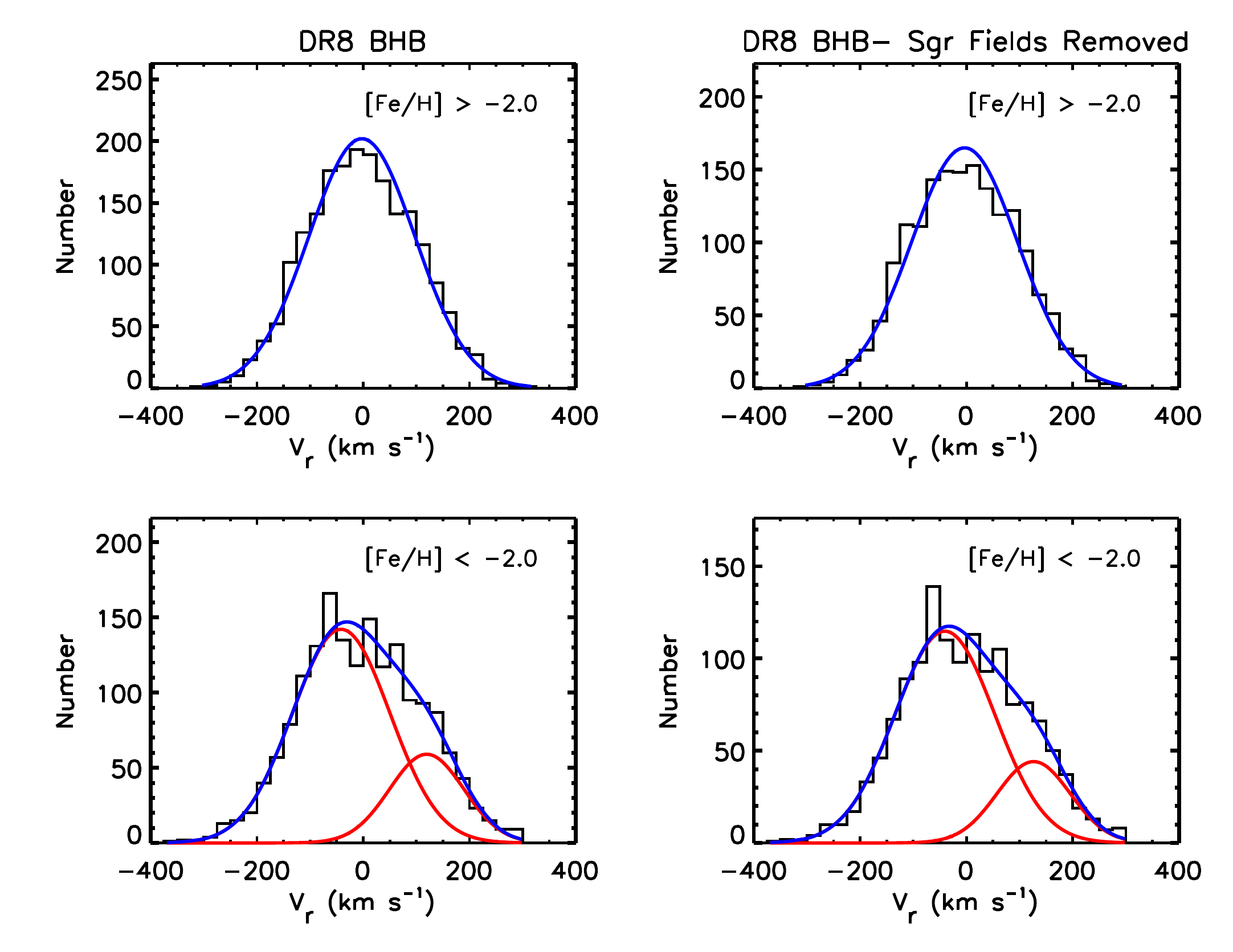}
\caption{{\it Left Panels:} Distribution of Galactocentric radial velocities for
a sample of well-classified Blue Horizontal-Branch (BHB) stars, based on the
sample from SDSS DR8 of Xue et al. (2011), including stars in the range of
Galactocentric distance 5 $< r < 40$ kpc, and with \z\ $> 4$ kpc. In the top
panel, the blue line is the best-fit Gaussian for stars with [Fe/H] $> -2.0$. 
In the bottom panel, for stars with [Fe/H] $< -2.0$, the red lines
represent components of the best two-component fit suggested by R-Mix, and the
blue line is the resulting mixture model. {\it Right Panels:} Similar, but for
the case where BHB stars from plug-plates in the direction of the two most
prominent wraps of the Sagittarius tidal stream have been removed. The
metallicity estimates are based on those derived using the procedures described
by Wilhelm, Beers, \& Gray (1999), which are optimal for these warmer stars
(\teff\ $ > 7000$~K).}
\label{fig:bhb_velgal}
\end{figure*}

The left-hand column of panels shown in Fig.~\ref{fig:bhb_velgal} compares the
distributions of Galactocentric radial velocities for two subsamples of the BHB
stars in the range $5 < r < 40$ kpc (which includes roughly 90\% of the full
sample reported by Xue et al. 2011), after removal of stars with \z\ $< 4$ kpc
(to ensure elimination of thick-disk BHB stars). The top panel, which applies to
BHB stars with [Fe/H] $< -2.0$, shows the best-fit Gaussian (obtained from the
R-Mix procedure), with a derived mean of $-15 \pm 2$ \kms\ and a dispersion of
100 $\pm 2$ \kms. R-Mix cannot reject the single-component hypothesis for this
subsample. As is immediately clear from inspection of the bottom panel, the
distribution of Galactocentric radial velocities for the BHB stars with [Fe/H]
$< -2.0$ exhibits rather different behavior. We emphasize that this subsample is
chosen from stars populating the same spatial distribution and having similar
distances (the median distance of the [Fe/H] $> -2.0$ subsample is 17.5 kpc; the
median distance of the [Fe/H] $< -2.0$ subsample is 18.8 kpc); only the
metallicity cut differs. Although the R-Mix procedure strongly rejects the
single-component hypothesis (with $p = 0.002$), the derived best-fit mean would
be $-16 \pm 3$ \kms, with a dispersion of 112 $\pm 2$ \kms. In order to obtain
an acceptable description of these data, R-Mix requires a two-component fit
(with means of I: $-54 \pm 13$ \kms, II: $108 \pm 18$ \kms, and dispersions of
I: 92 $\pm 6$ \kms, II: 70 $\pm 7$ \kms, respectively). 

The region of the Galaxy explored by the BHB stars considered here includes
possible members of the Sagittarius tidal stream (see Ruhland et al. 2011), so
we have carried out the same experiment as above, but with all BHB stars from
plug-plates in the directions toward the two most prominent wraps of the Sgr
stream removed from the analysis. The results are shown in the right-hand column
of panels in Fig.~\ref{fig:bhb_velgal}. Although the total numbers of BHB stars
are reduced, little else changes. The median distance of the [Fe/H] $> -2.0$
subsample is 16.8 kpc, while the median distance of the [Fe/H] $< -2.0$
subsample is 18.4 kpc, similar to the previous case. The best-fit Gaussian for
the [Fe/H] $> -2.0$ subsample, obtained from the R-Mix procedure, has a mean of
$-16 \pm 2$ \kms\ and a dispersion of 99 $\pm 2$ \kms. R-Mix cannot reject the
single-component hypothesis for this subsample. In the case of the [Fe/H] $<
-2.0$ subsample, the R-Mix procedure once again rejects the single-component
hypothesis (with $p = 0.02$); the derived best-fit mean would be $-16 \pm 3$
\kms, with a dispersion of 112 $\pm 2$ \kms. To obtain an acceptable description
of these data, R-Mix requires a two-component fit (with means of I: $-52 \pm
13$ \kms, II: $114 \pm 19$ \kms, and dispersions of I: 94 $\pm 6$ \kms, II: 69
$\pm 8$ \kms, respectively). 
  
In both cases (with or without the Sgr fields included) a two-sample K-S test
rejects the hypothesis that the subsamples of stars split at [Fe/H] $= -2.0$ are drawn
from the same parent population at high statistical significance ($p = 0.01$ for
the first instance and $p = 0.03$ in the second instance). Similarly, a
parametric F-test that the dispersions are the same rejects this hypothesis in
both cases, with $p < 0.001$. The observed radial velocities and
metallicities of the BHB stars are telling us that the halo is not a single
population.

\subsection{Additional Evidence (4):\\  The Metallicity Distribution Function of
SDSS DR8 BHB Stars\\ and Variation with Galactocentric Distance}

We now reconsider evidence similar to that presented in C07, which reported
an apparent variation of the nature of the MDF for horizontal-branch stars
selected from SDSS DR5.  Here we make use of the same sample discussed above,
the BHB stars from Xue et al. (2011), which is substantially larger.  As before,
we have considered this sample for two instances -- with and without inclusion
of BHB stars from plug-plates in the directions of the Sgr tidal stream.

\begin{figure*}
\centering
\includegraphics[angle=90,scale=0.65]{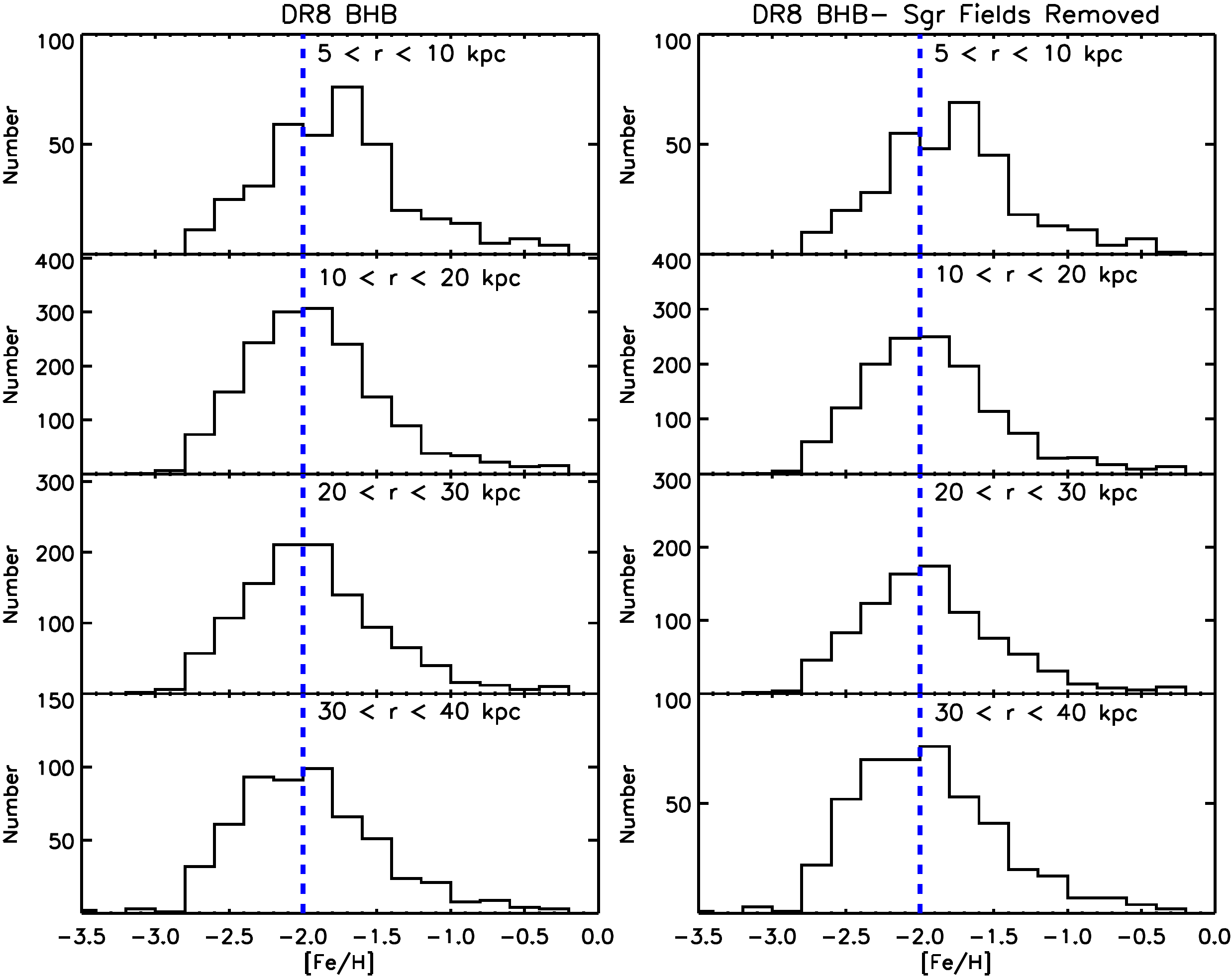}
\caption{{\it Left Panels:} As-observed MDF for the Xue et al. (2011) BHB stars
for various cuts on the distance from the Galactic center, $r$. Stars with \z $<
4$ kpc have been removed from the sample. {\it Right Panels:} Similar, but for
the case where BHB stars from plug-plates in the direction of the two most
prominent wraps of the Sagittarius tidal stream have been removed. In both
cases, the nature of the MDF appears to shift from the top panels, which exhibit
distributions that we associate with the inner-halo population, over to
distributions in the lower three panels that are dominated by
the outer-halo population. The metallicity estimates are based on those derived
using the procedures described by Wilhelm, Beers, \& Gray (1999), which are
optimal for these warmer stars (\teff\ $ > 7000$~K). The dashed blue line
provides a reference at [Fe/H]$ = -2.0$.}
\label{fig:bhb_mdf_zdist}
\end{figure*}

The left-hand column of panels in Fig.~\ref{fig:bhb_mdf_zdist} shows the
distribution of [Fe/H] in intervals of Galactocentric distance for the full
sample, after removal of stars with \z\ $< 4$ kpc. The peak of the MDF in this
panel is at [Fe/H] $ = -1.7$, close to what we would associate with dominance by
an inner-halo population. Comparing with the lower panels, the peak of the MDF
shifts to [Fe/H] $\sim -2$, close to that we would associate with an outer-halo
population. The strength of the low-metallicity tail in the lower panels is also
clearly greater than seen in the top panel. The fractions of stars with [Fe/H]
$< -2.0$ increase from 31\% for stars with 5 $< r <$ 10 kpc to between 46\% and
49\% for stars at larger Galactocentric distances. Indeed, a K-S test of the
null hypothesis that the MDFs of stars shown in the lower panels for the
individual cuts on Galactocentric distance $r$ could be drawn from the same
parent population as the stars shown in the top panel, against an alternative
that the stars are drawn from more metal-poor parent MDFs, is rejected at high
levels of statistical significance (one-sided probabilities of $p < 0.001$ for
all three higher cuts on Galactocentric distance).

The right-hand column of panels in Fig.~\ref{fig:bhb_mdf_zdist} is similar, but
with the BHB stars from plug-plates in the directions of the Sgr tidal stream
removed. As can be verified by inspection, little changes. The fractions of
stars with [Fe/H] $< -2.0$ increase from 30\% for stars with 5 $< r <$ 10 kpc to
between 44\% and 48\% for stars at larger Galactocentric distances. A K-S test of the
null hypothesis that the MDFs of stars shown in the lower panels for the
individual cuts on Galactocentric distance $r$ could be drawn from the same
parent population as the stars shown in the top panel, against an alternative
that the stars are drawn from more metal-poor parent MDFs, is rejected at high
levels of statistical significance (one-sided probabilities of $p < 0.001$ for
all three higher cuts on Galactocentric distance), as in the previous case. 

It is also interesting that the most dramatic shift in the appearance of
the MDFs in both the left-hand and right-hand columns of panels shown in
Fig.~\ref{fig:bhb_mdf_zdist} occurs between the top panels at $5 < r < 10$
kpc, and the next larger cuts in distance, at $10 < r < 20$ kpc, and hardly
changes thereafter. Indeed, a K-S test of the third distance cuts compared
with the second cuts, as well as for the fourth cuts compared with the
third cuts, cannot reject the hypothesis that the samples are drawn from
the same parent populations. Such a behavior might be easier to understand
as a superposition of multiple populations, with different mean metallicities,
rather than by invoking a continuous change that might be expected if a
strong metallicity gradient were present in the halo of the Galaxy. In any
event, this behavior is difficult to reconcile with the hypothesis of a
single-halo population possessing a spatially invariant MDF.

\section{Further Evidence for the Dual Halo of the Milky Way}

Quite independent of the above discussion of the C07 and C10 calibration-star
samples and the DR8 BHB sample, a substantial amount of evidence in support of
the dual-halo interpretation has already appeared in the literature, or has
been recently submitted for publication. Below we summarize a few of the
examples we consider the most persuasive.

\subsection{From Inside the SDSS}

\subsubsection{The SDSS DR7 BHB Sample Analyzed by Deason et al. (2011)}

Deason et al. (2011) have examined a sample of BHB stars selected from SDSS
DR7, using a combination of color cuts and \logg\ and \teff\ intervals from
the SSPP. Their sample overlaps substantially with that used by Xue et al.
(2008) to obtain estimates of the mass and constraints on the mass profile
of the Milky Way, which was based on SDSS DR6 (Adelman-McCarthy et al.
2008). The Deason et al. sample is larger (by about a factor of two) than
that used by Xue et al., not only due to the additional targets that were
included in DR7, but also because their selection is not as restrictive.

Among other results, these authors have used a set of adopted distribution
functions to model the observed Galactocentric radial velocities as a
function of distance and metallicity. They found that this sample of halo
stars exhibits a dichotomy between a prograde, comparatively metal-rich
component ([Fe/H] $> -2$), and a retrograde, comparatively metal-poor
([Fe/H] $< -2$) component. Although these properties are quite similar to
those advocated by C07 and C10, they concluded that the existence of a
low-metallicity retrograde population may simply indicate that estimates of
the rotation of the Local Standard of Rest (LSR), for which they adopt the
IAU recommended value of 220 \kms, may be underestimated by some 20 \kms.
They also point out that their results contrast somewhat with those from
C07 and C10, in that {\it both} their retrograde and prograde populations
are found in the distant regions of the halo, and are not necessarily due
to a shift in stellar populations with distance from the Galactic center,
as envisioned by the Carollo et al. studies. While such details demand
further investigation, it is clear that the Deason et al. results would not
support a single-halo interpretation of the present data.

\subsubsection{Spatial Variations in the Metallicity and Density Profiles\\ for
Modeled Halo Components from de Jong et al. (2010)}

During the course of the SEGUE subsurvey conducted during SDSS-II (Yanny et
al. 2009), ten ``vertical'' (in Galactic coordinates) photometric scans of
width 2.5$^\circ$, crossing the Galactic plane at fixed longitudes, were
imaged in the $ugriz$ passbands. The purpose of these scans was to extend
previous SDSS imaging to include selected areas in the latitude range
$-50^{\circ} < b < +50^{\circ}$, and thereby obtain more detailed
information on the transition from the halo system to the disk system of
the Milky Way. In their analysis of these data, de Jong et al. (2010)
employed a CMD fitting approach, based on templates of old stellar
populations with differing metallicities, to obtain a sparse
three-dimensional map of the stellar distribution at $|$Z$| >$ 1 kpc.

The maps of de Jong et al. (2010) provide clear {\it in situ} evidence for
a shift in the mean metallicity of the Milky Way's stellar halo -- within
$r \lesssim 15$ kpc their derived stellar halo exhibited a mean metallicity
of [Fe/H] $\sim -1.6$, changing to [Fe/H] $\sim -2.2$ at larger
Galactocentric distances. In addition, inspection of the spatial-density
profiles of their template populations (their Fig. 7) suggested rather
different behaviors for their ``inner-halo like'' template population and
that of their ``outer-halo like'' template population. Their derived
inner-halo density profile falls off rapidly with distance from the
Galactic center to $r \sim 15-20$ kpc; beyond this region a substantially
lower density, slowly varying, outer-halo density profile was found. Note
that the de Jong et al. analysis was restricted to distances $r < 30$ kpc.
When a single power-law was fit to this entire region they obtained an
index of $n = -2.75 \pm 0.07$, in excellent agreement with the previous
work of Bell et al. (2008) and Juri\'c et al. (2008).

Clearly, these findings provide compelling support for the kinematics-based
inferences of C07 and C10, as confirmed by our own reanalysis above, as
well as by the newly-considered BHB samples.

\subsubsection{Rejection of Single Power-Law Descriptions of the Milky Way's
Halo\\ based on Deep Repeated SDSS Imaging from Watkins et al. (2009) and Sesar et al. (2010)}

The region known as Stripe 82 (an area of $\sim$250 deg$^2$ along the
Celestial Equator) has been multiply scanned in the $ugriz$ filters over
the course of SDSS and its extensions, in particular during the Supernova
Survey conducted as part of SDSS-II (Frieman et al. 2008).

Both Watkins et al. (2009) and Sesar et al. (2010) have argued persuasively
that single power-law profiles are incapable of describing the spatial
variation of the halo system. These authors presented evidence, based on
both RR Lyrae stars and main-sequence stars, that the halo stellar
number-density profile significantly steepens beyond a Galactocentric
distance of $r \sim$30 kpc. It is worth noting that a ``steepening''
density profile might also be envisaged as describing the behavior of a
profile that suffers a large drop in stellar number density at a given
distance, and goes over to a more slowly varying profile with distance, as
was seen in the de Jong et al. (2010) analysis.

\subsubsection{The Identification of Spatial Autocorrelation in [Fe/H]
based on ECHOS\\ from Schlaufman et al. (2011)}

Schlaufman et al. (2009) have described the results of a systematic,
statistical search for elements of kinematically-cold halo substructure
(ECHOS) amoung the inner-halo metal-poor main-sequence turnoff (MPMSTO)
population identified during the course of SDSS/SEGUE. A by-product of the
search for ECHOS described by these authors is a catalog of MPMSTO stars
far more than 4 kpc from the Galactic plane, between 10 and 17.5 kpc from
the Galactic center, and free of both surface-brightness and
radial-velocity substructure, which they refer to as a `pure smooth halo
sample.' In the second paper in this series (Schlaufman et al. 2011a), they
analyzed co-added MPMSTO spectra to derive the average [Fe/H] and
[$\alpha$/Fe] for ECHOS, as well as for the smooth component of the halo
along the same line of sight as each ECHOS. They reported that the MPMSTO
stars in ECHOS were systematically more metal rich and less [$\alpha$/Fe]
enhanced than the MPMSTO stars in the smooth component of the halo,
concluding that the chemical-abundance pattern of ECHOS was best matched by
that of a massive dSph galaxy with $M_{tot} \gtrsim 10^9 M_{\odot}$. 

In the third paper of the series, Schlaufman et al. (2011b) quantify the
degree of spatial chemical inhomogeneity and spatial variation in chemical
abundance in the smooth component of the halo, using their
substructure-cleaned sample of MPMSTO stars. These authors report that the
classical smooth halo component ceases to be the dominant component of the
stellar population of the halo system beyond about 15 kpc from the Galactic
center, and furthermore, that there exists significant spatial coherence in
[Fe/H] in the MPMSTO population beyond this distance. They suggest from
these findings that the relative contribution of disrupted low-mass
galaxies to the stellar population of the smooth halo increases with
radius, becoming observable relative to the classical kinematically smooth
halo beyond 15 kpc. They also find that the morphology of the halo system
in the [Fe/H]/[$\alpha$/Fe] plane inside of 15 kpc is not well-matched by
phased-mixed tidal debris. Instead, they argue that the smooth halo inside
of 15 kpc is likely formed through a combination of {\it in situ} star formation
and dissipative major mergers at high redshift.  They conclude that their
results are ... ``consistent with the dual halo idea advanced in Carollo et
al. (2007,2010)...''.

\subsection{From Outside the SDSS}

Over the past several decades there have been numerous studies of the
nature of the halo system that provide evidence indicating the halo of the
Milky Way may not comprise a single population, based on analyses of the
spatial number-density profiles of halo tracer objects (such as globular
clusters or field horizontal-branch stars), and of the kinematics of small
subsamples of these. Here we briefly mention a subset of these, based on
data obtained outside the SDSS.

Representative spatial-analysis papers include Hartwick (1987),
Sommer-Larsen \& Zhen (1990; using density profiles inferred from local
kinematics), Preston et al. (1991), Zinn (1993), Kinman et al. (1994), and
Chiba \& Beers (2000) (using density profiles inferred from local
kinematics), all of which reached similar conclusions. According to these
studies, the halo is best described as flattened in the inner regions, but
going over to a much more spherical distribution at larger radii. Similar
work has been conducted with ever increasing sample sizes in recent years.
Examples include analyses of RR Lyraes based on the data from the QUEST
survey (Vivas \& Zinn 2006), as well as from the LONEOS sample (Miceli et
al. 2008). In this latter example, Miceli et al. argued for the presence of
a dual halo in order to account for the apparently very different spatial
profiles of Oosterhoff Type I and Oosterhoff Type II variables in their
sample. Most recently, Sesar et al. (2011) used deep imaging data from the
CFHT Legacy Survey to study the distribution of near-turnoff main-sequence
stars in the Galactic halo along four lines of sight, to heliocentric
distances of $\sim$35 kpc. They found that the halo stellar number-density
profile becomes steeper at Galactocentric distances greater than $r \sim
30$ kpc, and emphasize that single power-law models are strongly disfavored
by the data.

Representative kinematical-analysis papers include Norris \& Ryan (1989),
Allen et al. (1991), Carney et al. (1996), Wilhelm et al. (1996), Borkova
\& Marsakov (2003), Kinman et al. (2007), and de Propris et al. (2010).
Several of these papers (Norris \& Ryan 1989; Allen et al. 1991; Carney et
al. 1996) emphasized the clear presence of individual stars, in particular
those with low metallicities, associated with large retrograde motions.
Below [Fe/H] $= -2.0$, the numbers of retrograde stars in both Allen et al.
(1991) and Carney et al. (1996) are well in excess of the numbers expected
for fair samples drawn from a single halo population with little or no net
rotation. Indeed, both sets of authors commented on the need for a complex
halo in order to accommodate their observations. Wilhelm et al. (1996),
Borkova \& Marsakov (2003), and Kinman et al. (2007) all reported
significant net retrograde motions for subsamples drawn from their BHB and
RR Lyrae star samples, respectively. Most recently, de Propris et al.
(2010) have presented radial velocity data from BHB stars indicating that
the velocity dispersion profile of the halo appears to increase towards
large Galactocentric radii, while the stellar velocity distribution is
non-Gaussian beyond 60 kpc. They concluded that the outer halo consists of
a multitude of low luminosity overlapping tidal streams from recently
accreted objects.

\section{Summary and Conclusions}

We have considered the criticisms of S10 in detail, and demonstrated that
their claim that the retrograde signature of the outer-halo population is
due to incorrect distance determinations or improper assignments of the
stellar luminosity classes appears spurious. The original assertions of C07
and C10 are in fact confirmed by our analysis. The distance scale advocated
by S10 was based on their adoption of the incorrect main-sequence absolute
magnitude relationship from the work of I08. We have shown that, for redder
stars, this scale is roughly 10\% (in the median distance) shorter than the
correct globular cluster-based scale suggested by I08, increasing to 18\%
shorter for bluer stars near the main-sequence turnoff, independent of
metallicity. Comparison with a calibrated isochrone approach by A11
indicates that, for stars with [Fe/H] $< -2.0$ (which dominate the
membership of the outer-halo population), the S10 scale is roughly 10\%
shorter for redder stars and 17\% shorter for bluer stars near the
main-sequence turnoff. The distance scale for main-sequence dwarfs with
[Fe/H] $< -2.0$, based on the revised C10 classifications (including
reassigned TO $\rightarrow$ D stars), agree with the determinations of both
I08 and A11 at a level of 6-10\%, which we consider more than adequate.

We have carried out an abbreviated kinematic analysis for the very
low-metallicity stars that dominate the proposed outer-halo component,
using the distance scales of the various studies considered above. Based on
this analysis, we confirm the existence of a significant retrograde
population (with a large velocity dispersion), which C07 and C10 associated
with this structure. Furthermore, we have shown that the origin of the
retrograde signature at low metallicity is traceable to the asymmetric
distribution of the observed proper motions, not to the assignment of
incorrect distances. The shift of the MDF for the C10 sample with distance
from the Galactic plane, the distribution of Galactocentric radial
velocities for BHB stars from SDSS DR8, as well as the variation of the BHB
MDF with Galactocentric distance, in addition to other evidence from inside
and outside the SDSS, are all consistent with a kinematically and/or
chemically distinct superposition of inner- and outer-halo populations in
the Milky Way, and not with a homogeneous single-halo population.

Over the span of the past quarter century, data from many independent
surveys, based on very different selection criteria, distance estimation
techniques, and analysis methodologies, have been pointing with ever
increasing confidence to the conclusion that a single-halo description is
no longer valid for the Milky Way. We have summarized the most relevant
results here, although much additional evidence exists. 

It is important to note that recent high-resolution, cosmologically-based
simulations, in particular those of Zolotov et al. (2009, 2010), Font et
al. (2011), and Tissera et al. (2011), now include at least approximate
prescriptions for the star formation and dissipative accretion processes,
as well as other pertinent baryonic physics such as metal-dependent cooling
and supernova feedback, greatly expanding their predictive power. This new
generation of simulations indicates that a dual halo (with different
stellar spatial-density profiles and clear metallicity shifts between an
inner- and outer-halo population) is {\it a generic expectation} for large
spiral galaxies such as the Milky Way and M31.  The Font et al. (2011) analysis
emphasizes the structural characteristics of their simulations; they find
remarkable similarities to the observed characteristics of the Milky Way
reported by C07, C10, and de Jong et al. (2010), including the detection of
``breaks'' in the stellar spatial-density profiles and metallicity drops of
from 0.6 to 0.9 dex in going from the regions they associate with an inner
halo to an outer halo. The Tissera et al. (2011) study emphasizes the contrasting
chemical properties of the components they associate with an inner halo and
outer halo in their suite of simulations, separated on the basis of the
binding energy of the stellar particles. As shown in their Fig. 4, clear
shifts are seen in the mean [Fe/H] of the particles in their inner/outer
haloes, always in the sense that the outer halo is substantially more
metal-poor than the inner halo. The median differences they obtain between
these components range over $-0.4 < \Delta$[Fe/H] $< -0.9$, well-bracketing
the difference in the peak metallicities of the Milky Way's inner halo and
outer halo reported by C07 and C10 ($\Delta $[Fe/H]$ = -0.6$). Finally, the
McCarthy et al. (2011) analysis of the kinematical behavior of star
particles from the Font et al. (2011) simulations appears completely
consistent with what might have been expected from the observational
analysis of C07 and C10. We draw particular attention to their Fig. 10,
which illustrates clear differences in the (velocity-dispersion scaled)
rotational properties of their `{\it in situ} spheroid' (inner halo) and
their `accreted spheroid' (outer halo). Their Fig. 11, a representation of
a Toomre diagram for their simulations, clearly indicates that their {\it
in situ} spheroid and accreted spheroid possess strongly contrasting
velocity dispersions, as also has been shown by C07 and C10 to be an
observed characteristic of the halo system of the Milky Way.

Additional analyses of the SDSS/SEGUE stars, beyond those in the
calibration-star subset, should help to strengthen the observational case
for (at least) a dual halo, and to refine estimates of the parameters that
describe the individual components. Ultimately, geometric distances from
Gaia for stars in the halo populations will eliminate any remaining
questions concerning the impact of uncertain photometric parallaxes on
these conclusions. However, our view is that presently available data
already reject the single-halo interpretation beyond reasonable doubt. 

\acknowledgements

Funding for SDSS-I and SDSS-II has been provided by the Alfred P. Sloan
Foundation, the Participating Institutions, the National Science Foundation, the
U.S. Department of Energy, the National Aeronautics and Space Administration,
the Japanese Monbukagakusho, the Max Planck Society, and the Higher Education
Funding Council for England. The SDSS Web Site is http://www.sdss.org/.

The SDSS is managed by the Astrophysical Research Consortium for the
Participating Institutions. The Participating Institutions are the American
Museum of Natural History, Astrophysical Institute Potsdam, University of Basel,
University of Cambridge, Case Western Reserve University, University of Chicago,
Drexel University, Fermilab, the Institute for Advanced Study, the Japan
Participation Group, Johns Hopkins University, the Joint Institute for Nuclear
Astrophysics, the Kavli Institute for Particle Astrophysics and Cosmology, the
Korean Scientist Group, the Chinese Academy of Sciences (LAMOST), Los Alamos
National Laboratory, the Max-Planck-Institute for Astronomy (MPIA), the
Max-Planck-Institute for Astrophysics (MPA), New Mexico State University, Ohio
State University, University of Pittsburgh, University of Portsmouth, Princeton
University, the United States Naval Observatory, and the University of
Washington.

TCB and YSL acknowledge partial support from grants PHY 02-16783 and PHY
08-22648: Physics Frontiers Center/Joint Institute for Nuclear Astrophysics
(JINA), awarded by the U.S. National Science Foundation. MC acknowledges support
from a Grant-in-Aid for Scientific Research (20340039) of the Ministry of
Education, Culture, Sports, Science and Technology in Japan. Studies at ANU of
the most metal-poor populations of the Milky Way are supported by Australian
Research Council grants DP0663562 and DP0984924. ZI acknowledges support from
NSF grants AST 06-15991 and AST 07-07901, as well as from grant AST 05-51161 to
LSST for design and development activities. DA acknowledges support
provided by the National Research Foundation of Korea to the Center for
Galaxy Evolution Research.  

{\it Facilities:} \facility{SDSS}.

\vfill\eject

\appendix

\section{Comments on the S11 Paper}

As noted in the Introduction, we now attempt to re-establish some linearity
in the progress of the discussion of the C07/C10 papers (the original
presentation of the case for the presence of a dual halo for the Milky
Way), the S10 criticism of these works (provided as 1012.0842v1 on the
arXiv preprint server), the original Beers et al. (2011; B11) response to
S10 (provided as 1104.2513v1 on the arXiv preprint server), the published
version of the \Sch\ (2011; S11) paper, and the revised version of the B11
submission (this paper). To accomplish this, all of the remarks presented
in the main body of the present paper have been confined to discussion of
the S10 manuscript. Here, we offer a few responses to the issues that were
raised by S11.

The S11 paper concludes in their abstract that ``Finally, we note that
their revised analysis presented in Beers et al. does not alleviate our
main concerns." This statement is the result of S11's puzzling choice to
ignore the great majority of points made by B11, including all of the
substantial evidence for the presence of a dual halo {\it beyond those
arising} from the C07/C10 analysis, which has been supplemented with
additional information in Section 5 of the present paper. Furthermore, much
of the S11 paper concentrates on the effect that the presence of
misclassified TO stars has on the conclusions of C07/C10, even though B11
presented a set of procedures for their identification and reassignment,
and demonstrated that the presence of an outer halo with properties similar
to those originally claimed remains intact, even after the reassignments
have been performed.  

The S11 paper claims that the main-sequence relationship they adopted from
I08 (Eqn. (A1) in I08, modified by S10 in ways that serve to bring it into
closer agreement with Eqn. (A7) from I08 that B11 argue is the more
appropriate one) is actually a reasonably accurate description of the
zero-age main sequence (ZAMS) in the required color range at low
metallicity. Equation (A1) was never intended to describe the ZAMS -- this
relationship was obtained as an intermediate step in the I08 development of
their recommended absolute-magnitude relationship, Eqn. (A7). The S11 paper
now claims that they find no significant differences between their analysis
carried out with the modified version of Eqn. (A1) and that of Eqn. (A7),
while we have shown that rather large differences can emerge (Fig. 4 of
B11). Perhaps their modifications of Eqn. (A1) have corrected some of this
discrepancy, but caution must still be exercised. The discussion of S11
puts a great deal of faith in their use of the BaSTI isochrones
(Pietrinferni et al. 2004, 2006), using them as the basis for the statement
in their conclusions that "... we have made use of direct isochone
distances, which fully corroborate our findings." This statement implies
that the I08 Eqn. (A7) relationship is somehow defective with respect to
their choice, even though B11 demonstrate excellent agreement between the
I08 Eqn. (A7) relationship and the set of {\it calibrated isochrones}
described by A11 (Fig. 8 of B11). No mention is made in S11 of the B11
consideration of the A11 treatment.

The S11 paper (as did the S10 draft before it) makes use of the metallicity
estimates from SDSS DR7, without application of the correction procedure
recommended by C10, as if this correction were not needed. In fact, as
several authors of C10 and B11 were responsible for assigment of the
metallicities for DR7, they recognized that such a correction was required
on the basis of comparison with the available high-resolution spectroscopic
study of numerous low-metallicity stars. A recent verification can be found
in the slides presented by Aoki (2011) at The Third Subaru International
Conference on Galactic Archaeology\footnote{{\tt
http://www.naoj.org/SubaruConf11/slides/subaru3\_aokiwako.pdf}}; further
discussion is presented by Aoki et al. (2011, in preparation). 

The S11 paper (and the S10 draft) calls attention to the possibility of the
presence of a Lutz \& Keller (1973) bias in distance estimates, which may
give rise in turn to asymmetries in the derived azimuthal velocities.
Even though a Lutz \& Keller bias may exist, it cannot be responsible for
the entire retrograde signature, as discussed in Section 5.1 above.
Indeed, S11 neglects to address the objection noted in B11 (and expanded
upon on the present paper) that consideration of the proper-motion
components alone, for the stars identified as belonging to the highly-
retrograde tail of the outer-halo population, are clearly not drawn from the
same parent population as other stars of similarly low metallicity (Fig. 12
of B11).
 
In their conclusions, S11 make the assertion that the revised outer-halo
mean velocity has changed from $-158$ \kms\ (pointing to Table 1 from C10)
to $-59 \pm 20$ \kms\ in Table 2 of B11. In fact, this apparent
disagreement is spurious. Table 1 of C10 does not present the final result
of C10, but rather gives results for a robust clustering analysis of
\vphi, which applies to all stars satisfying the listed cuts, including the
misclassified TO stars, and does not make use of any of the models or
analysis methods of C10.  Table 2 of B11 refers to an R-Mix analysis of the
spectroscopically assigned D stars; the resulting net motion is in fact
statistically consistent with that obtained by the maximum-likelihood
result for the outer-halo component provided in Table 5 of C10, $-80 \pm
13$ \kms; the error bars -- derived in different ways -- overlap at the
one-sigma level. The level of agreement with the C10 result is apparently
even better if one specifies to the lower portion of Table 2 of B11, which
uses the (larger) full set of D, TO, and SG/G stars: $-94 \pm 23$ \kms.  

Finally, in their conclusions, S11 have mischaracterized the result of the
luminosity reclassifications of the TO stars carried out by B11, claiming
``... moving a considerable fraction of the wrongly identified TO stars up
to the subgiant/giant branch will make the distance overestimate for the
misidentified dwarfs among them even more severe". This statement is made
in spite of the fact that, in Section 2.2 of B11, it is made clear that
85\% of TO stars with \feh\ $< -2.0$ are reassigned to D status, while
those reassigned to SG/G status only comprise 14\%. The impression given to
the reader by S11 would appear to be that just the opposite has occurred.

\vfill\eject

\vfill\eject

\begin{deluxetable}{cccc}
\tablewidth{0pt}
\tablenum{1}
\tablecaption{Luminosity Class Refinements for Main-Sequence Turnoff Stars}
\tablehead{
\colhead{Former Class} &
\colhead{\teff\ Range } &
\colhead{Gravity Interval } &
\colhead{New Class}
}
\startdata
    TO     &     $\geq$ \tcrit\ &  3.75 $\leq \log g < 4.00$   &       TO  \\
    TO     &     $\geq$ \tcrit\ &  3.50 $\leq \log g < 3.75$   &       TO  \\
    TO     &     $<$ \tcrit\    &  3.75 $\leq \log g < 4.00$   &       D   \\
    TO     &     $<$ \tcrit\    &  3.50 $\leq \log g < 3.75$   &      SG/G
\enddata
\end{deluxetable}
\vfill\eject

\begin{deluxetable}{ccccccccccc}
\tablewidth{0pt}
\tablenum{2}
\tablecolumns{11}
\tablecaption{R-Mix Results for the Low-Metallicity Subsample:  Kinematic Parameters}
\tablehead{
\colhead{Sample} &
\colhead{Number} &
\colhead{$\langle$V$_{\phi,I}$$\rangle$} &
\colhead{} &
\colhead{$\sigma_{V_{\phi,I}}$} &
\colhead{} &
\colhead{$\langle$V$_{\phi,II}$$\rangle$} &
\colhead{} &
\colhead{$\sigma_{V_{\phi,II}}$} &
\colhead{} &
\colhead{\it p-value }\\
\colhead{} &
\colhead{} &
\colhead{(km~s$^{-1})$ } &
\colhead{} &
\colhead{(km~s$^{-1})$ } &
\colhead{} &
\colhead{(km~s$^{-1})$ } &
\colhead{} &
\colhead{(km~s$^{-1})$ } &
\colhead{} &
\colhead{1-Comp} \\
}
\startdata
\cutinhead{Spectroscopically Identified Dwarfs}
\multicolumn{11}{c}{[Fe/H] $< -$2.0 ; $Z_{\rm max} >$ 0 kpc} \\
            &      &                  &  &              &  &             &  &               &   & \\
Rev. C10    & 1298 & $-$77  $\pm$ 57  &  & 117 $\pm$ 15 &  & 44 $\pm$ 11 &  &79 $\pm$ 10    &   & $< 0.001$\\
I08         &  635 & $-$84  $\pm$ 66  &  & 94  $\pm$ 21 &  & 53 $\pm$ 20 &  &72 $\pm$ 8     &   & $< 0.001$\\
A11         &  360 &  $-$100 $\pm$ 28 &  & 124 $\pm$ 11 &  & 52 $\pm$ 12 &  & 70 $\pm$ 8    &   & $< 0.001$\\
S10         &  694 &  $-$46  $\pm$ 47 &  & 85  $\pm$ 11 &  & 72 $\pm$ 14 &  &64 $\pm$ 8     &   & $< 0.001$\\
            &   &                  &  &              &  &             &  &               &   & \\
\multicolumn{11}{c}{[Fe/H] $< -$2.0 ; $Z_{\rm max} >$ 5 kpc} \\
            &   &                  &  &              &  &             &  &               &   & \\
Rev. C10    &  469 & $-$59  $\pm$ 20  &  & 147 $\pm$ 11 &  & 8 $\pm$ 10  &  &78 $\pm$ 11    &   & $< 0.001$\\
I08         &  184 & $-$200  $\pm$ 40 &  & 83  $\pm$ 28 &  & 20 $\pm$ 8  &  &84 $\pm$ 6     &   & $< 0.001$\\
A11         &  173 &  $-$395 $\pm$ 15 &  & 35 $\pm$ 12  &  & $-$24 $\pm$ 12&  & 116 $\pm$ 9 &   & $< 0.001$\\
S10         &  119 &  13 $\pm$ 7      &  & 92 $\pm$ 5   &  &  \nodata    &  &  \nodata      &   & $0.8$\\
\cutinhead{All Stars -- D, TO, and SG/G}
            &      &                  &  &              &  &             &  &               &   & \\
\multicolumn{11}{c}{[Fe/H] $< -$2.0 ; $Z_{\rm max} >$ 0 kpc} \\
            &      &                  &  &              &  &             &  &               &   & \\
Rev. C10    & 1471 &  $-$91 $\pm$ 23    &  & 124 $\pm$ 8  &  & 40 $\pm$ 9  &  &80 $\pm$ 7     &   &  $< 0.001$\\
            &      &                  &  &              &  &             &  &               &   & \\
\multicolumn{11}{c}{[Fe/H] $< -$2.0 ; $Z_{\rm max} >$ 5 kpc} \\
            &      &                  &  &              &  &             &  &               &   & \\
Rev. C10    &  577 &  $-$94 $\pm$ 23    &  & 153 $\pm$ 9   &  & 12 $\pm$ 10 &  & 83 $\pm$ 10   &   &  $< 0.001$\\

\enddata
\end{deluxetable}
\newpage

\newpage

\newpage

\newpage

\newpage

\newpage

\newpage

\newpage

\newpage

\newpage

\newpage

\newpage

\newpage

\newpage

\newpage

\end{document}